\documentclass[12pt]{article} 
\usepackage[hyperfootnotes=true]{hyperref}
\hypersetup{
  colorlinks=true,        % false: boxed links; true: colored links
  linkcolor=blue,         % color of internal links
  citecolor=cyan,         % color of links to bibliography
}
\usepackage{epsfig}
\usepackage{float}
\usepackage{amsmath}
\usepackage{mathtools,nccmath}
\usepackage{amssymb}
\usepackage{amsfonts}
\usepackage{latexsym}
\usepackage{graphicx}
\usepackage{xcolor}
\usepackage{cite}
\usepackage[T1]{fontenc}
\usepackage[utf8]{inputenc}
\newlength{\abstractwidth}
\usepackage{ wasysym } 
\usepackage{mciteplus} 
\usepackage{subcaption}
\usepackage{bbold}
\usepackage{slashed}

\usepackage{marvosym}

\usepackage{mathbbol}
\DeclareSymbolFontAlphabet{\amsmathbb}{AMSb}%

\usepackage{amsthm}
\usepackage{hyperref}
\usepackage{color}
\usepackage{mathrsfs}
\usepackage{soul}

\usepackage[outline]{contour}
\newcommand*{\fancy}[1]{{\color{white}\contour{black}{#1}}}

\DeclareMathAlphabet{\mathpzc}{OT1}{pzc}{m}{it}

\usepackage{calligra}
\DeclareMathAlphabet{\mathcalligra}{T1}{calligra}{m}{n}
\DeclareFontShape{T1}{calligra}{m}{n}{<->s*[2.2]callig15}{}
\newcommand{\scripty}[1]{\ensuremath{\mathcalligra{#1}}}

\def\be {\begin{equation}}
\def\ee {\end{equation}}
\def\bea {\begin{eqnarray}}
\def\eea {\end{eqnarray}}
\def\bc {\begin{center}}
\def\ec {\end{center}}
\def\bfg {\begin{figure}}
\def\efg {\end{figure}}
\def\bi {\begin{itemize}}
\def\ei {\end{itemize}}

\def\sq {\sqrt}

\DeclareMathOperator{\Tr}{Tr}

\def\beq{\begin{equation}}
\def\eeq{\end{equation}}
\def\br{\begin{eqnarray}}
\def\er{\end{eqnarray}}
\newcommand{\eel}[1] {\label{#1}\end{equation}}

\usepackage{authblk}
\usepackage[symbol*]{footmisc}

\DefineFNsymbolsTM{otherfnsymbols}{%
  \Bat  \mathsection
  \MoveUp   \mathsection
}%

\setfnsymbol{otherfnsymbols}

\newenvironment{arabicfootnotes}
  {\par\edef\savedfootnotenumber{\number\value{footnote}}
   
   \setcounter{footnote}{0}}
  {\par\setcounter{footnote}{\savedfootnotenumber}}

\begin{document}

\title{Einstein's Equations and the pseudo--Entropy of pseudo--Riemannian Information Manifolds}

\author{Hassan Alshal
\thanks{halshal@scu.edu}}

\affil{\textit{Department of Physics, Santa Clara University,\linebreak 500 El Camino Real, Santa Clara, CA 95053, United States}}

\date{}

\maketitle

\begin{abstract}

Motivated by the corrected form of the entropy--area law, and with the help of von Neumann entropy of quantum matter, we construct an emergent spacetime by the virtue of the geometric language of statistical information manifolds. We discuss the link between Wald and Jacobson approaches of thermodynamic/gravity correspondence and Fisher \textit{pseudo}--Riemannian metric of information manifold. We derive in detail Einstein's field equations in statistical information geometric forms. This results in finding a quantum origin of a positive cosmological constant that is founded on Fisher metric. This cosmological constant resembles those found in Lovelock's theories in a de Sitter background as a result of using the complex extension of spacetime and the Gaussian exponential families of probability distributions, and we find a time varying dynamical gravitational constant as a function of Fisher metric together with the corresponding Ryu--Takayanagi formula of such system. Consequently, we obtain a dynamical equation for the entropy in information manifold using Liouville--von Neumann equation from the Hamiltonian of the system. This Hamiltonian is suggested to be non--Hermitian, which corroborates the approaches that relate non--unitary conformal field theories to information manifolds. This provides some insights on resolving ``the problem of time''.

\medskip
\noindent
\end{abstract}

\newpage

  %  \starttext \baselineskip=17.63pt \setcounter{footnote}{0}
   \tableofcontents

\begin{arabicfootnotes}
 
\section{Introduction}

One of the main challenges in physics is to find a fundamental dynamics between geometry/gravity and quantum matter. And many approaches such as string theory, gravity/field correspondence, and loop quantum gravity try to tackle this problem in different ways and frameworks; check \cite{Rovelli:1997qj} for detailed review. In this paper, we approach the problem from the perspective of information manifolds and entropy. In the last few years, information geometry has earned a great interest in fields like machine learning \cite{amari2010information} and deep learning in physics \cite{baldi2014searching}. The information manifold and entropy concepts, particularly the \emph{relative entropy}, are extremely useful in understanding many physical patterns that include, but not limited to, quantum computers \cite{vedral2002role}, chemistry\cite{shell2008relative}, biological systems \cite{baez2016relative} and even economy \cite{avellaneda1998minimum}. The entropy--area law, corrected and generalized by the outside von Neumann entropy \cite{Bekenstein:1972tm,Bekenstein:1973ur,Bekenstein:1974ax}, is introduced to the information geometry in order to check that geometry obeys the second law of thermodynamics and preserves information \cite{Maldacena:2020ady}. Our proposal to approach the gravity/quantum problem arises intuitively from looking at the information paradox in black hole physics from the entanglement entropy perspective \cite{Ryu:2006bv,Hubeny:2007xt} along with the formally established the second law of thermodynamics for black holes Noetherian charges \cite{Wald:1993nt,Iyer:1994ys,Wall:2011hj} and the quantum origin of spacetime and Einstein equations \cite{Jacobson:1995ab}. 

The outline of  this work is organized as follows. Right after this passage and within the introduction section, we summarize the reasons behind correlating the holographic principle with relevant entropy, \emph{coarse--grained entropy}, Kullback--Leibler (KL) divergence, and Fisher information metric. Then in section (\ref{EntropyPreservedInfo}), we investigate the corrected entropy--area law by von Neumann entropy in order to both satisfy the second law of thermodynamics and preserve information. We compute a form of corrected entropy--area law in the light of Liouville--von Neumann equation so that we get a dynamical relation between the quantum Hamiltonian and the time variation of both entropy and area. Later in the same section, we briefly review the gravity/thermodynamics correspondence, developed by Jacobson \cite{Jacobson:1995ab} and by Wald \cite{Wald:1993nt}, since we will use that later in the following sections. We comment on the relation between the quantum origin of spacetime and the dynamical equation of expansion rate. In section (\ref{quantuminformation}), we discuss the importance of von Neumann entropy and its maximized coarse--grained version \cite{Almheiri:2020cfm}, and we provide a detailed study of the geometric representation of the density matrix related to both entropies. We cover the essential structures of the information geometry, and we apply the thermodynamics/geometry correspondence to derive an information geometry form of Einstein field equation. The Fisher information metric will be introduced to measure how far the \textit{cumulant probability density functions} are away from each others after being varied with respect to the microscopic variables. In other words, the Fisher metric tells us about the quota of information the microscopic variables carry in the statistical manifolds \cite{Amari85,Balian:1986jrj}. Thus, we study the correlations between variables using the cumulant probability density functions as good candidates to count on, instead of the moments of distributions, while developing the Fisher metric. For example, think of the mean square error as a second moment of the error. When it is differentiated with respect to the microscopic value, the varied cumulant probability density measures the level of proficiency in the model--data fitting.\footnote{In the statistical sciences jargon it is called \textit{score} or \textit{informant} of the likelihood function \cite{Schervish}.} This also appears in KL divergence that measures the difference between probability distributions. The last concept plays a fundamental role in relating Fisher metric to Shannon/von Neumann entropy, which is what we show later. And that concept can be seen mathematically as a second derivative with respect to the microscopic variables\footnote{In the statistical sciences jargon it is called \textit{observed information}, and is equal to the mean square error of the same second derivative applied on the cumulant probability distribution.} acting on the KL divergence, which is nothing but the inverse of the probability distribution of Shannon/von Neumann entropy. Such structure is very similar to the mathematical definitions of curvatures in Riemannian manifolds, which is addressed also in the same section. Thus, Fisher metric is in fact a metric from which the curvature structure in the information manifold is obtained \cite{Amari2016} once proved to be endowed with other properties of Riemannian manifolds. This will lead to reformulate Einstein field equations in the corresponding information manifold. Additionally, a positive cosmological term in the Einstein equations is obtained in the information manifold. Later, we relate the components of the field equations to commulant functions and get a more detailed informatic description to the gravitational constant. In section (\ref{EntropyManifold}), we introduce a dynamical equation of entropy in the information manifold using only quantum information geometry without using any classical components. It is a new combination of von Neumann master equation, the von Neumann entropy, and the gravitational entropy formula, and its generalization to statistical manifolds. Then, we use the RT formula for the statistical manifold to describe its corresponding \emph{pseudo--entropy} \cite{Doi:2022iyj,Doi:2023zaf}. In section (\ref{conclusion}), we discuss and comment on the findings.

The so--far achieved developments in entropic information theory have led to conceptually rich ideas like entanglement entropy \cite{Eisert:2008ur}. This line of thoughts can be traced back to the discovered relation between the area and the black hole entropy in the Bekenstein--Hawking \cite{Bekenstein:1973ur,Hawking:1975vcx}. The holographic principle, developed by 't Hooft \cite{tHooft:1993dmi} and Susskind \cite{Susskind:1994vu}, suggests finding a correspondence between the $3D$ volume and the $2D$ area, which leads to the known gauge/gravity correspondence, or the AdS/CFT as a quantum field theory with local degrees of freedom \cite{Maldacena:1997re}, and the sufficiently described entropy at the \emph{microscopic level} in contrary to Bekenstein--Hawking law. Thus, we are allowed to address the entanglement entropy   \cite{Bombelli:1986rw,Srednicki:1993im}, which describes the quantum information load in the quantum states, as a stored information encoded on the geometric features of the space \cite{Harlow:2014yka}. But since different surface geometries are indeed associated with different entanglement entropies, the area in Bekenstein--Hawking law is suggested to be replaced by another area law for the \textit{extremal surfaces}, known as Ryu--Takayanagi (RT) surfaces in holography models \cite{Ryu:2006bv,Hubeny:2007xt}, which provides a clue to rethink of the entropy built of microscopic variables as the fundamental underpinnings of the spacetime classical geometry upon reintroducing RT surfaces to Wald's formula for the entropy of black holes \cite{Dong:2013qoa}. We can exploit RT law as the commoner understanding of the law allows using it to describe the von Neumann entropy of gravity--coupled quantum systems as a \emph{pseudo--entropy}, controlled by the system's microscopic variables, in dS space \cite{Doi:2022iyj,Narayan:2022afv,Doi:2023zaf,Cotler:2023xku} and in AdS space \cite{Nakata:2020luh,He:2023eap,Chen:2023gnh,Chu:2023zah}. Pseudo--entropy is a property for holographic systems obtained from generalizing the entanglement entropy using the transition matrix of two different states living in a black hole/environment system. In contrary to the usual definition of entropy, pseudo--entropy is allowed to take negative or complex values; more on its properties can be found int Ref. \cite{Nakata:2020luh,Mollabashi:2020yie}. Aside from our approach, it is worth referring to the yet--to--be--revealed relation between pseudo--entropy and the so--called \emph{tensor network} \cite{Mori:2022xec}. There are numerous statistical gravitational tensor network models based on RT entanglement approach \cite{Chirco:2017vhs,Chirco:2019dlx,Chirco:2019gsd,Colafranceschi:2021acz}. It is argued that semiclassical coarse--graining of holographic states, as realized in \textit{tensor networks}, result in a flow in spacetimes approaching RT surface \cite{Murdia}. Moreover, random tensor network is proposed as a statistical gravitational model of a massive scalar field on a thermally fluctuating background using neural network algorithm, which recovers the background geometry fluctuation from multi-region RT entanglement entropy \cite{Lam:2021ugb}.

The complicated relations among the microscopic variables, together with the difficulties associated with measuring those variables, are the motives behind why it is always more convenient to express the statistical phenomena corresponding to the microscopic variables in terms of the stochastic variables that are ruled by more fundamental and relatively easier--to--measure laws. Such reductionism in the description automatically will lead to select some variables to be \textit{relevant} and other microscopic variables to be \textit{irrelevant} \cite{Balian:1986jrj}.  Relevant variables are all parameterized as functions in time or any other affine parameter. We can think of the relation between the microscopic and macroscopic variables like the relation between the speed of gas molecules and the temperature of the whole sample, both can be used to describe different types of thermodynamical energies. These microscopic relevant variables should be averaged, using the relevant density function, such that they define the macroscopic or the stochastic variables. And the averaging process could be done using any density matrix as we discuss in details in subsection (\ref{vec.sp.construct}). Additionally, any two different densities, constructed from those relevant variables, are favored over each other according to which of them is capable of defining maximum entropy that corresponds to least amount of data loss due to the unfavorable unavoidable effects coming from irrelevant information \cite{Jaynes:1978}. Such density is known as the \textit{canonical coarse--grained density}, and its corresponding entropy is known as the \textit{relevant entropy}; it is the candidate entropy to reformulate the RT law of statistical manifolds as discussed in section (\ref{quantuminformation}). 

Due to the mathematical difficulties a person might encounter while trying to find the exact states and their corresponding densities, it is suggested to replace the von Neumann general relevant entropy with other more specified entropies: the \emph{coarse--grained relative entropy}, where this entropy is always larger or equal to the fine--grained von Neumann entropy, and we comment on that in section (\ref{vec.sp.construct}). Yet we emphasize that the coarse--grained entropy is ``lossy but true'' entropy \cite{Flak} as it depends on the macroscopic variables of the system, see Ref. \cite{Kozlov} for more on that. Also, more on the relation between the fine--grained entropy and von Neumann entropy can be found in Ref. \cite{Almheiri:2020cfm}. It is argued that one can coarse--grain any type of quantum entropy, such as entanglement entropy, using \textit{observational entropy} \cite{Safranek}. The technique of coarse--graining obtained with help of the projectors acting on Hilbert space has been previously used in information geometry\footnote{Balian \textit{et al.} \cite{Balian:1986jrj} did not restrict the physical observables to be the Hilbert space projectors but Safr\'anek \emph{et al.} did \cite{Safranek}.} \cite{Balian:1986jrj}. It is worth noting that there are some generalized versions of quantum entropies, such as Rényi and Tsallis entropies, that do not manifestly take the form of the von Neumann entropy. Yet they can be treated as von Neumann entropy upon applying the proper limits \cite{AManko:2016dwe}. Notice also that observational entropy is a quantum analogue of the classical Boltzmann entropy. More importantly, observational entropy shows the measurement limitations when one tries to get more precise information even if the density state is more precisely known. Thus, the process of coarse--graining is inevitable because even pure states span over more than one macrostate in the phase space due to superposition. Yet, observational entropy is bounded from below by von Neumann entropy, and equals to the later if the former satisfies the coarse--graining conditions in Ref. \cite{Safranek}, i.e. after several consecutive coarse--graining processes, one can end up having a fine--grained entropy. Generally for finite--dimensional systems, the observational entropy can be expressed as a \textit{relative entropy}. Statistically speaking, observational entropy can take the form of KL divergence, which is a type of relative entropy, from which we obtain Fisher information metric in information manifolds.

\section{Information and Spacetime Thermodynamics}\label{EntropyPreservedInfo}
 In a stationary black hole with perfect fluid environment, the black hole entropy follows Bekenstein-Hawking formula, while the rest of the entropy is allocated to the perfect fluid \cite{Bekenstein:1972tm,Bekenstein:1973ur,Bekenstein:1974ax,Bardeen:1973gs}. However, the entropy can be generalized to include the interactions, especially the scattering processes, of the external entangled quanta and Hawking radiation \cite{Bombelli:1986rw,Almheiri:2020cfm,Maldacena:2020ady}. In order to obtain an entropy--area law that respects the second law of thermodynamics and preserves information, we need the full entropy of a black hole to contain both the entropy of what is inside the horizon and the entropy of the quantum matter outside the horizon \cite{Maldacena:2020ady,Penington:2019npb, Almheiri:2019psf}. This means the generalized entropy law \cite{Bekenstein:1972tm,Bekenstein:1973ur,Bekenstein:1974ax} would take the form
\be\label{SBH}
S_{\text{gen}}= \frac{A_{\text{H}}}{4 G \hbar}+ S_{\text{matter}},
\ee
where $S_{\text{gen}}$ is the generalized entropy of the black hole, $A_{\text{H}}$ is the area of black hole horizon that could be coarse--grained, and $S_{\text{matter}}$ is the von Neumann entanglement entropy of the matter outside the black hole. The constants are Planck constant $\hbar$ and Newton's gravitational constant $G$.\\ On one side, von Neumann entropy for a quantum-mechanical system described by a density matrix $\rho$ is given by
\be\label{SVN}
S_{\text{matter}}= -\Tr(\rho \ln \rho).
\ee
On the other side, the time-evolution equation of the density matrix $\rho$ is given by Liouville--von Neumann equation \cite{Neumann},
\be\label{LVN}
\frac{d \rho}{dt} = \frac{1}{i\hbar}[H,\rho],
\ee
\\
where $H$ is the Hamiltonian of the considered quantum system. Since the trace operator commutes with the differential time operator, we can in principle write a time evolution equation for the von Neumann entropy. For that purpose, we use the straightforward mathematical trick
\be\label{trick}
\frac{d \rho}{d t}= \frac{d}{d t}(\rho \ln \rho)- \frac{d \rho}{d t} \ln \rho .
\ee
\\
Using Eq. (\ref{trick}) in Eq. (\ref{LVN}), we get the quantum time evolution equation as follows
\be\label{trick1}
i \hbar \frac{d}{d t}(\rho \ln \rho)- i \hbar \frac{d\rho}{d t} \ln \rho= [H,\rho] .
\ee
\\
We take the trace of both sides and use the fact that trace operator commutes with the differential time operator. Then, we substitute  Eq. (\ref{SVN}) in Eq. (\ref{trick1}) to get
\be\label{tVN}
-i \hbar ~\frac{d }{d t}S_{\text{matter}}= \Tr\bigg[i \hbar \frac{d \rho}{d t} \ln \rho+ [H,\rho]\bigg] .
\ee 
\\
Thus, an equation for the time evolution of von Neumann entropy is obtained. For a black hole system, the von Neumann entropy satisfies the second law of thermodynamics only through introducing the fine--grained entropy as we mentioned in the introduction. The coarse--grained entropy is obtained by tracing every possible density matrix using the environment observables \cite{Almheiri:2020cfm}. Then, the fined--grained entropy is maximized to become coarse--grained one. Thus, the semiclassical approximation guarantees that the von Neumann entropy is that of the quanta outside the black hole in curved spacetime. Consequently, we use Eq. (\ref{SBH}) to rewrite the time evolution of entropy as follows
\be\label{result}
-i \hbar \frac{d S_{\text{gen}}}{d t}+ i \frac{1}{4 G} \frac{d A_{\text{H}}}{d t}
=\Tr\bigg[i \hbar \frac{d\rho}{d t} \ln \rho+ [H,\rho]\bigg].
\ee
\\
The last equation introduces a a relation between black hole full entropy, black hole horizon area, and density matrix of quantum states. As we observe here, we did not make any additional assumptions to get to Eq. (\ref{result}). It appears that Eq. (\ref{result}) is a quantum/semi--classical form of entropy-area law for the black hole. It introduces a relation between geometry (Area) and quantum matter (Density Matrix). We will reconsider this relation after we connect entropy as a macroscopic quantum quantity to the area as a geometric quantity. But before that, it is worth noting that the two concepts are related in general within the \textit{non--dissipative} systems, i.e. systems with $dS_{\text{gen}}/dt=0$. This assumption is valid as $S_{\text{gen}}$ is the \textit{total} corrected entropy of the black hole, and when $A_{\text{H}}$ and the RT extremal surface $A_{\text{ext.RT}}$ coincide or an ``island'' is realized \cite{Hubeny:2007xt,Faulkner:2013ana}. Such extremal surface realization is related to the existence of flat plateau in the corresponding Page curve \cite{Gautason:2020tmk,Cao:2021ujs} after minimizing the generalized entropy of the extremal surfaces.\footnote{Generalized extremal entropy: $S_{\text{gen.RT}}=min\left[A_{\text{ext.RT}}/4\hbar G+S_{matter}\right]$, which looks similar to Eq. (\ref{SBH}) \cite{Almheiri:2020cfm}. Minimizing the entropy recovers the Page curve as required and guarantees that the fine grained entropy will not exceed the coarse grained one as emphasized in the next section.} Therefore, Eq. (\ref{result}) becomes
\be\label{fine-grained}
i \frac{1}{4 G} \frac{d A_{\text{H}}}{dt}
=\Tr\bigg[i \hbar \frac{d \rho}{d t} \ln \rho+ [H,\rho]\bigg].
\ee
When $A_{\text{H}}$ and $A_{\text{ext.RT}}$ coincide, the von Neumann term becomes negligible, and the area term dominates the entropy, hence the importance of Eq. (\ref{fine-grained}). Upon solving the previous equation, we have two possible explanations. Either the area would have the fine--grained entropic definition, which is expected as the horizon contains the information of the entangled particles that fall inside it \cite{Bianchi:2012ev}, or the fine--grained quantum part of the black hole entropy would have a coarse--grained geometric meaning based on the microscopic variables of the density, the entropy, and the manifold. Before we discuss the second meaning, which is what we do in section (\ref{EntropyManifold}), we emphasize that all previous equations are derived assuming that Liouvillian mechanics stems from $d\rho/dt=0$. Also, the Hamilton--Jacobi equation shows that $H=-~\partial\mathcal{A}/\partial t$, where $\mathcal{A}$ is the action. As we will see later, this Hamiltonian, and consequently the action, depends on a cumulant function that will be treated as a scalar field to obtain Einstein equations in the information manifold. Wald \cite{Wald:1993nt} noticed that all what one needs to do is to express the entropy as a function of the density of the state then applies Liouvillian mechanics to solve the Hamilton--Jacobi equation in order to generate the action that will be extremized to get the conserved quantities together with Euler--Lagrange equations. This is why we swiftly review the gravity/thermodynamics correspondence developed by Jacobson \cite{Jacobson:1993vj,Jacobson:1995ab} based on Wald approach \cite{Wald:1993nt}.

Assuming Rindler frame of references, Jacobson found that the Einstein equations can be obtained from the entropy--horizon area relation together with the laws of thermodynamics \cite{Jacobson:1995ab}. According to Unruh's radiation \cite{Unruh:1976db}, the radiation temperature detected by a Rindler observer is directly proportional to the uniform acceleration $``a"$ of that observer. Applying the equivalence principle to Rindler transformations \cite{Rindler:1960zz,Rindler:1966zz} guarantees that the vacuum state in the generic spacetime is locally more or less the same as the Minkowski vacuum, i.e. the time involved in defining the quantum operators in vacuum, measured by the accelerated observer, should be close to that measured by an observed in a flat spacetime \cite{Chirco:2009dc}. Thus, each point in the spacetime has its own local Rindler horizons, both past and future, with Killing fields in the null directions of the horizons. The relations between heat flux and the black hole hairs, i.e. the mass and the angular momentum,\footnote{Electromagnetic energy is better represented separately inside the density matrix $\rho$.} are discussed in \cite{Bardeen:1973gs,Unruh:1983ms,Wald:1993nt,Wald:1999wa} through the Hamiltonian formulation of the first law of thermodynamics. In brief, the heat flow of such system could be defined to obey the averaged null energy condition along a time--like geodesic
\begin{equation}\label{QTkk}
    \delta Q =-a \int T_{\mu\nu} k^{\mu}k^{\nu} tdt\delta A,
\end{equation}
where $T_{\mu\nu}$ is the energy momentum tensor, $k^{\mu}$ is the null vector, $\delta A= \sqrt{\gamma}dA$ is the differential element in the congruence cross sectional area of the horizon, and $\sqrt{\gamma}$ is the determinant of the induced metric of the spatial area element of the horizon element $dA$. The last equation stems originally from Wald's formula \cite{Jacobson:1993vj,Jacobson:1995uq}. Also, the last equation is well founded in flat spacetime and independent of the Hamiltonian formulation of the first law of thermodynamics \cite{Baccetti:2013ica}. But the Hamiltonian formulation can also be used to obtain Eq. (\ref{QTkk}) with careful considerations upon applying it on the Rindler wedge, e.g., in the holographic approach the considered Hamiltonian should be \emph{modular} \cite{Faulkner:2013ica,Ruiter:2018iuq}. Now for the null geodesic of $k^{\mu}$ that generates the horizon,\footnote{Higher order contributions obtained from the shear terms should not be ignored in nonequilibrium thermodynamical systems \cite{Chirco:2009dc}.} the Raychaudhuri equation gives
\begin{equation}\label{theta}
    \theta=-tR_{\mu\nu}k^{\mu}k^{\nu}.
\end{equation}
where the affine parameter $\theta$ is the rate of change of $\sq{\gamma}$ \cite{Poisson:2011nh}, i.e.
\begin{equation}\label{theta.gamma}
    \theta=\frac{1}{\sq{\gamma}}\frac{d}{dt}\left(\sq{\gamma}\right),
\end{equation}
which also describes the expansion of $\delta A$.
The last two equations can be combined to give
\begin{equation}\label{thetaRA}
    \theta A\frac{d\theta}{dA}=-R_{\mu\nu}k^{\mu}k^{\nu}.
\end{equation}
As in \cite{Jacobson:1995ab}, Eq. (\ref{QTkk}) and Eq. (\ref{theta}) yield the relation
\begin{equation}\label{Q/A}
    \frac{\delta Q}{\delta A}= \frac{a\hbar}{8\pi\ell^{2}_{P}},
\end{equation}
which demands that the Einstein equations become
\begin{equation}\label{Einstein}
    G_{\mu\nu}=\frac{8\pi\ell^{2}_{P}}{\hbar}T_{\mu\nu}~,
\end{equation}
where the $G_{\mu\nu}$ is the Einstein tensor and $\ell_{P}$ is the Planck length.

It is worth mentioning that another approach of understanding gravity as an entropic force is suggested by Verlinde in \cite{Verlinde:2010hp,Verlinde:2016toy} but we will not review it here. Rather, one might say that nothing new in this section. But Eq. (\ref{thetaRA}) and Eq. (\ref{Q/A}) entice us to consider the existence of a new relation between $dS=\delta Q/T$ and $\delta\theta$ at the microscopic level of $\rho$ considering the von Neumann definition of entropy. Such $(S,\theta)$ relation is indeed a dynamical relation, and we will study that relation in the context of information geometry in section (\ref{EntropyManifold}). But first we need to study the meaning of the geometric quantities $G_{\mu\nu}$ and $\theta$ in the classical spacetime using Fisher Riemannian metric of information manifold. In the next section, we adopt the Fisher metric as it guarantees extending the geometric quantities to any quantum theory. Moreover, the observational entropy is realized as the KL divergence, which is another information quantity from which we can obtain the Fisher metric.

\section{Entropy and Riemannian Geometry} \label{quantuminformation}

With help of \textit{Lie group thermodynamics} in the framework of the covariant formalization of geometrized thermodynamics \cite{Souriau:1978,Barbaresco}, we can expand the analysis that resulted in Eq. (\ref{fine-grained}) in the speculative direction of the geometric interpretation of the fine--grained quantum part of the black hole entropy. Balian \emph{et al.} gauge theory of thermodynamics \cite{Balian:1986jrj}, as an extension of Fisher Riemannian metric of information manifold, guarantees extending the geometric interpretation to any quantum theory. Later, we introduce Balian \emph{et al.} Riemann metric of the density matrix space as the Hessian of the fine--grained entropy. This metric stems directly from the endeavors of Ruppeiner information metric \cite{Ruppeiner:1995zz}. The metric is situated as a canonical function in between the space of states and the space of observables. This involves Legendre transforms just like those in Liouvillian mechanics. We also comment on the similarities between the Hessian structure in our approach and that of gravity in the framework of superstrings \cite{Gross:1986iv}.

\subsection{Vector space construction}\label{vec.sp.construct}
According to Balian \emph{et al.} \cite{Balian:1986jrj}, it is useful to focus only on the space of the density states as the density state is more suitable for information gathering in comparison with observables themselves. But when it comes to the density state space metric, we would rather use the fine--grained entropy to define such metric. This choice is made as the density state could concede an \textit{incomplete description} for the information relative to the observables, i.e. finding the average value of any observable $\hat{\mathcal{O}}$ using the usual relation
\begin{equation}\label{aver}
    \langle \hat{\mathcal{O}} \rangle = \text{Tr}\left[\hat{\mathcal{O}} ~\hat{\rho}\right]
\end{equation}
does not define a unique $\rho$ as there are different $\rho$'s that sufficiently define this value. Notice that the fine--grained entropy described in Eq. (\ref{SVN}) has a \textit{global maximum} as we infer from Fig. \hyperref[fig1]{(1)}. Then, all densities satisfying Eq. (\ref{aver}) give equivalent entropies, and hence, the loss in the information of the observables due to the different $\rho$'s is irrelevant. But the density $\rho_0$ corresponding to maximum entropy is more favorable to describe the density space metric as, by construction, it has the minimum information to calculate different types of observables. Even for the irrelevant information, $\rho_0$ contains the least of them, and thus it corresponds to the entropy that ``\emph{quantifies our ignorance about the precise quantum state of the system}'' \cite{Almheiri:2020cfm}. This is why the relevant von Neumann entropy $S[\rho_0]$ is that of the macroscopic thermal phenomena such as the thermodynamically gravitational quantities in the dissipative systems. To elaborate the above discussion, we focus in Eq. (\ref{aver}) on the $\hat{\rho}\hat{\mathcal{O}}$ components. The density $\hat{\rho}_{\hat{\mathcal{O}}}$ is chosen to be the density proportional to the eigenprojectors of $\hat{\mathcal{O}}$ in the corresponding sub--Hilbert space. Therefore, all the off--diagonal elements in the general $\hat{\rho}$ will be disregarded, and the left information is stored in the $\hat{\rho}_{\hat{\mathcal{O}}}$. Since $\hat{\rho_0}$ is the density that has all the relevant information of all observables similar to $\hat{\mathcal{O}}$, it is clear that $S[\rho_0]>S[\rho]$. Additionally, $S[\rho]_{coarse}\geqslant S[\rho_0]$, where $S[\rho]_{coarse}$ is the coarse--grained entropy that provides a measure to $S_{\text{gen}}$ and could match fine--grained later when the previously mentioned inequality becomes a direct equal relation between the two types of entropy \cite{Almheiri:2020cfm}.

Now we construct a geometric interpretation of the previously discussed thoughts such that the ``\emph{variation of the relevant entropy represents a transfer of information between the relevant and the irrelevant variables}'' and ``\emph{dissipation appears as a leakage of the relevant information}'' \cite{Balian:1986jrj}. With the help of the algebraic structure endowed in the observable space, the geometric construction in the language of manifolds\footnote{In that sense, the density matrix $\rho$ should be seen as a \emph{pre--probability}, see eq. (15.46) and section (15.4) of Ref. \cite{griffiths_2001}.} would help us in explaining Eq. (\ref{result}--\ref{fine-grained}) in a purely quantum information way. The observable $\hat{\mathcal{O}}$ has eigenvectors $|\alpha^k\rangle$ in the space states, and therefore, it can be written as 
\begin{equation}
    \langle\hat{\mathcal{O}}\rangle=\sum\limits_{k}\langle \alpha^k|\rho|\alpha^k\rangle\text{Tr}\Big[ |\alpha^k\rangle\langle \alpha^k|\mathcal{O}\Big].
\end{equation}
To define the vector space of the observable $\hat{\mathcal{O}}$, we set the components and the bases as
\begin{subequations}\label{vO}
   \begin{align}
   \mathfrak{o}_{\mu}&:= \langle \alpha^k|\rho| \alpha^k\rangle,\\
   f^{\mu}&:=|\alpha^k\rangle\langle\alpha^k|\mathcal{O},
   \end{align}
\end{subequations}
respectively as the components and the bases of the \textit{Liouville vector} $\vec{\mathcal{O}}=\mathfrak{o}_{\mu}f^{\mu}$.

As the set of all observables $\hat{\mathcal{O}}$ is characterized by the $\rho$, and $\rho$ itself is an operator that can be written in terms of any set of orthonormal bases $\{|i\rangle\}$. Then, we can rewrite Eq. (\ref{aver}--\ref{vO}) such that we define the density operator as a vector $\vec{\rho}$ with components
\begin{equation}\label{Vrho}
    \rho^{\mu}:=\langle f^{\mu}\rangle=\text{Tr}\left[f^{\mu}\rho\right].
\end{equation}
Then, when $\mathcal{O}$ becomes $\rho$, Eq. (\ref{aver}) becomes
\begin{equation}\label{OVrho}
\overrightarrow{\langle \hat{\rho}\rangle}:=\vec{\rho}=e_{\mu}\rho^{\mu},
\end{equation}
where $\rho^{\mu}$ act as the components of the averaged observable $\langle\hat{\rho}\rangle$ in its \textit{Liouville vector representation} $\overrightarrow{\langle\hat{\rho}\rangle}$, and the basis components for such vector are defined as $e_{\mu}:=|i\rangle\langle i|$ that are dual to $f^{\mu}$, i.e. $e_{\mu}\cdot f^{\nu}=\delta_{\mu}^{~\nu}$. In such representation, Eq. (\ref{aver}) can be seen as a \textit{bilinear relation}
\begin{equation}\label{bilinear}
    ||\vec{\mathcal{O}}||=\langle \vec{\mathcal{O}},\vec{\rho}\rangle:=\mathfrak{o}_{\mu}\rho^{\mu}.
\end{equation}

To make the above construction more elaborate, let's say that $\hat{\mathcal{O}}$ commutes with the spatial observable $\hat{x}$ in a space with Euclidean signature, i.e. the $\vec{\mathcal{O}}$ is a function in the components of $\hat{x}$. Then, the density $\rho^{\mu}$ would correspond to the averaged components $\langle x^{\mu}\rangle$. As the density space is made of all the $\langle x^{\mu}\rangle$ including the irrelevant ones, this means it is valid to embed the function $\vec{\mathcal{O}}$ as a hypersurface in the space spanned by $\rho^{\mu}$ after being expressed in terms of the dual basis of $\vec{\rho}$. This surface is called \textit{the surface of reduced states}, and it is extremized at $\rho_0$. At the same time, it is much easier to describe the density in terms of the microscopic variables $\xi^{\mu}$ and the stochastic variables $\mathbb{X}^{\mu}$, both as functions parameterized by the spacetime observables $x^{\nu}$ or their averages, and we will do that later in subsection (\ref{FisherKLdiv}). Speaking of irrelevant averaged observables, relevant bases in are those of the density space needed for the calculation of mean values of specified observable, e.g. the metric, and can be isolated from the irrelevant bases. One can separate them using a special operator called the \emph{projection operator}, we will encounter this operator in subsection (\ref{Euclidstruct}). Conceptually, the relevant components of the $\hat{\rho}_{\hat{\mathcal{O}}}$ are those of $\rho_0$. But the time rate change of the relevant parts is indeed a function in some relevant and some irrelevant parts \cite{Gemmer:2009}. This is why we will upgrade Eq. (\ref{result}--\ref{fine-grained}) to become \emph{master equations} as we will see in section (\ref{EntropyManifold}). But for now, we just finished setting the stage up for the debut of the density space. Next, we show this space could be promoted to a density manifold.

\begin{figure}[H]\label{fig1}
\hspace*{-0.45in}
\centering
    \includegraphics[width=1.1\textwidth]{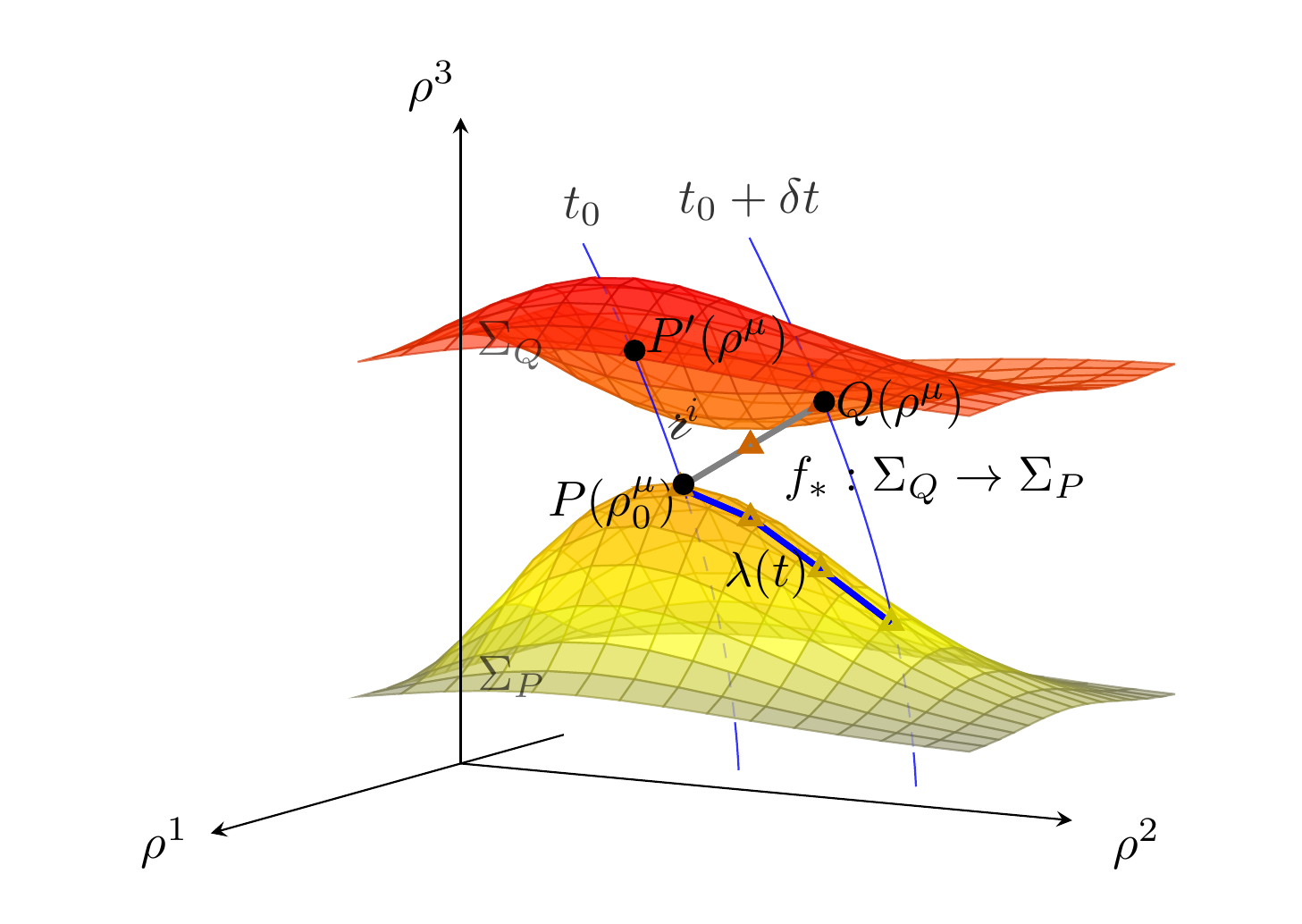}
    \caption{\small Density manifold and the evolution of entropy surfaces away from reduced state $\rho_0$.}
\end{figure}

\subsection{Density manifold}\label{DensityManifold}
In the density space, $\rho$ is guaranteed to be a function in time such that Eq. (\ref{LVN}) is satisfied. Geometrically this means that $\rho(t)$, as a point in such space, evolves from the point $P(\rho_0)$ on the surface $\Sigma_P$, where $S[\rho]=S[\rho_0]$, to another point on the same $\Sigma_P$ with $S[\rho]$ along some trajectory $\lambda$ \cite{Grabert:2006}. This is defined as an exponential function just like how we relate in Riemannian geometry a vector to the points belonging to the trajectory that the vector is tangent to. Thus
\begin{equation}\label{exp}
    \exp_P:t\cdot\rho^{\mu}\to\lambda_{\rho^{\mu}}(t).
\end{equation}
As $t\in[0,1]$, then $\exp(1.\rho^{\mu})=\lambda_{\rho^{\mu}}(1)$ is the final point $P(\rho)(t)$ at the trajectory. Meanwhile $\exp(0.\rho^{\mu})=\lambda_{\rho^{\mu}}(0)$ is the initial point $P(\rho_0)$. The $\rho^{\mu}_0$ can take any direction. Therefore, its $i^{th}$ component $\rho^{i}$ along the tangent of the trajectory is given by
\begin{equation}\label{tangent.rho}
\rho^{i}_{0}\equiv\frac{d\lambda(t)}{dt}\bigg|_{t=0}~ .
\end{equation}\\
Moreover, the $i^{th}$ component of $\rho^{i}_0$ is allowed to be in the direction of any irrelevant basis $\rho^{\mu}$ as long as it is tangent to the trajectory $\lambda$. This motivates considering the exponential function (\ref{exp}) as a diffeomorphism between the neighborhood of point $P$, which belongs to the state space, and the vector $\rho^{\mu}_0$, which belongs to the tangent space $\text{T}_P\Sigma$ at point $P$. If we normalize $\rho^{i}_0, \forall i$, then the orthonormal basis $\lbrace\rho^{\mu}_0\rbrace$ of $\text{T}_P\Sigma$ provides the isomorphism
\begin{equation}\label{isomorphTpSigma}
    E:\mathbb{R}^{n}\xrightarrow{\sim}\text{T}_P\Sigma,
\end{equation}
such that $E(\xi^1,\cdots,\xi^n)\vert_{P}=c_i\rho^{i}_0$.\\
The previous manifold--related definitions are valid regardless whether the surface is extremized on not. Therefore, there exists \textit{charts} $\psi$ on manifold $\mathcal{M}$ containing all $\lambda$'s and $\text{T}_Q\Sigma,$ for all $Q\in\mathcal{M}$, such that
\begin{equation}
    \psi:=(\exp\cdot E)^{-1}:\mathcal{M}\to\mathbb{R}^n.
\end{equation}

Also, the exponential map of the density manifold acts similarly to how the known the exponential map $\exp(tV)=\lambda_V(t)$ works between any Lie group and its Lie algebra, where $V\in$ the algebra, $\lambda_V(t)\in$ the group, and $t\in\mathbb{R}$. Then, there exists an analytic diffeomorphism in a neighborhood $U$ of $V=0$ such that, for the coordinates $\xi^{\mu}\in\mathbb{R}^n$ defined by the isomorphism $E$ in Eq. (\ref{isomorphTpSigma}), we find that $\exp(\xi_iV^i)\in\exp(U)$. This would define the necessary canonical chart.

Moreover, the entropy $S[\rho]$ never loses information as long as the initial point $P(\rho_0)$ evolves in a Hamiltonian trajectory. However, when we disregard the irrelevant information, we do something similar to the \textit{shift and lapse} such that the entropy $S[\rho]=S[\rho_0]$ evolves in time through the extemized and non-extremized surfaces, i.e. we can practice \textit{push-forward}, $f_*:\Sigma_Q\to\Sigma_P$, (or \textit{pull-back}) transformations between $\Sigma_P$ and $\Sigma_Q$, together with applying the rules of Lie derivatives in order to relate the surfaces to each others. Now we are ready to introduce a metric on this \textit{information manifold}.

\subsection{Manifold metric}
We notice that the exponent function in Eq. (\ref{exp}) transfers $\rho^{\mu}$ to a tangent space with same properties but at different point. Since the logarithmic function is the opposite to the exponent, then, in light of Eq. (\ref{bilinear}), we can safely say that $(\ln\rho)$ does the same job of $\rho$ except that the domain here becomes the \textit{dual tangent space}. This implies that $S[\rho]$ acts as a bilinear map between the density $\rho$ and the \emph{information content} $(-\ln\rho)$. Information content is a unit of information regarding some random variable and follows Shannon's definition of self--information. If the logarithm is to the base 2, the information content is called \emph{bit}. And for the natural logarithm, the Shannon information entropy coincides with the Gibbs entropy, which can be quantized to be the von Neumann entropy, and information content is called \emph{nit}. Information content plays a fundamental role in constructing the Fisher metric of information manifolds as we will see in subsection (\ref{FisherKLdiv}). As both $\rho$ and $(-\ln\rho)$ are unique geometric vectors, then $S[\rho]$ provides us with a linear isomorphic relationship between the tangents and the dual tangent spaces, i.e. the duality
\begin{equation}\label{duality}
    \rho\equiv\widetilde{(\ln\rho)}
\end{equation}
is legitimate. Therefore, there exists a map $\mathpzc{G}$ on the $\rho$ space such that
\begin{subequations}\label{G}
\begin{align}
    \mathpzc{G}&:(\ln\rho)_{\mu}\mapsto\mathpzc{G}_{\mu\nu}\rho^{\nu},\\
    \mathpzc{G}&:(\rho)^{\mu}\mapsto\mathpzc{G}^{\mu\nu}(\ln\rho)_{\nu}~.
\end{align}
\end{subequations}
The map $\mathpzc{G}$ is shown to be symmetric, real, and has positive eigenvalues in the Liouville vector representation \cite{Balian:1986jrj}. It is the best function to play the role of metic in the $\rho$ space. Then, we can use Eq. (\ref{G}) to demonstrate the entropy as the  bilinear function 
\begin{equation}\label{Sbilinear}
    S\langle\rho,\ln\rho\rangle:=-\mathpzc{G}_{\mu\nu}(\rho(\xi))\rho^{\mu}\rho^{\nu}.
\end{equation}
The $\rho$ space is dense enough--satisfying the topological features of manifolds--such that we can introduce the infinitesimal change $d\rho$. Therefore, Eq. (\ref{Sbilinear}) can be redefined infinitesimally such that the second differential in the entropy becomes the metric itself.\footnote{Such metric metric can be considered a Ruppeiner information metric \cite{Ruppeiner:1995zz}.} And the $\rho$ space becomes eligible for a promotion to be a Riemannian manifold endowed with the invariant distance
\begin{equation}\label{Qdistance}
    -ds^2:= d^2S\langle\rho,\ln\rho\rangle=-\mathpzc{G}_{\mu\nu}(\rho(\xi))d\rho^{\mu}d\rho^{\nu},
\end{equation}
where the map (\ref{G}) is explicitly defined as
\begin{equation}\label{Qmetric}
    \mathpzc{G}(e_{\mu},e_{\nu}):=\mathpzc{G}_{\mu\nu}(\rho(\xi))=-\frac{d^2S}{d\rho^{\mu}d\rho^{\nu}}.
\end{equation}
We find Eq. (\ref{Qmetric}) leads us to define the \textit{dual vector} $(-\ln\rho)$ as
\begin{subequations}
\begin{align}
-(\ln\rho) &:=\varrho_{\mu}f^{\mu} \label{ln.varrho},\\
\varrho_{\mu}&:=(-\ln\rho)_{\mu}\equiv\frac{\partial S}{\partial \rho^{\mu}}.
\end{align}
\end{subequations}
Notice that if the density is defined as usual $\rho=\exp(-\beta H)/Z$, it renders $\phi-\varrho=-\beta H$ as in Eq. (\ref{varrhoxE}), where $Z$ is the partition function, the energy $H$ is related to the entropy $S$ \cite{Iyer:1994ys,Oh:2017pkr} and is a function in the cumulant partition function $\phi$ that we use later to define the metric of the manifold. This notice encourages us to comment on the Hamiltonian operator in such system and study its relation to Eq. (\ref{result}--\ref{fine-grained}). So we introduce a \emph{hermitian entanglement Hamiltonian} to Eq. (\ref{result}-\ref{fine-grained}) as a \textit{superoperator} \footnote{The prefix \textit{super} has no Grassmann rings, i.e. has nothing to do with Supersymmetry or graded algebra in general.} $\mathcal{H}$ such that
\begin{subequations}\label{superH}
\begin{align}
&\mathcal{H}:=\mathcal{H}^{~\mu}_{\nu}e_{\mu}\otimes f^{\nu},\\
&\mathcal{H}^{~\nu}_{\mu}=\langle \mathcal{H} e_{\mu}, f^{\nu}\rangle.
\end{align}
\end{subequations}
In order to double check that the Hamiltonian in the information manifold is that of $\phi$, we follow Ref. \cite{2020AIPA}. Eq. (\ref{SVN}--\ref{tVN}) define the hermitian entanglement Hamiltonian similar to those in Ref. \cite{Bianchi:2012ev}. Additionally, the full Hilbert space of the black hole and the environment makes defines the state, and consequently the density, as that of the black hole and environment together \cite{Giddings:2021qas}. Thus, we define the Hamiltonian as $\mathcal{H}=\mathcal{H}_o+\mathcal{H}_i$, where $\mathcal{H}_o$ is the non-interaction Hamiltonian--mainly that of the black hole and that of the environment separately--and $\mathcal{H}_i$ is the interaction Hamiltonian that \emph{connects the black hole state to the environment throughout information exchange} \cite{Giddings:2021qas}. Now, Eq. (\ref{LVN}) does not restrict the Hamiltonian to be that of the quantum states alone or their interactions. This means that the entanglement entropy $S_{\text{matter}}$ has a piece that depends on $\mathcal{H}_i$. And the entropy corresponding to $\mathcal{H}_i$, that is responsible for the mutual information exchange between the black hole and the environment, is another von Neumann entropy, i.e. it depends on $\rho$, as a function of $\phi$, and other operators acting independently on the environment and the black hole \cite{Giddings:2017mym}. Consequently, The Hamiltonian in Eq. (\ref{result}--\ref{fine-grained}) is expected to be $\mathcal{H}_i$ as a function of $\phi$. This result will help us when we reconsider Eq. (\ref{result}--\ref{fine-grained}) in section (\ref{EntropyManifold}). For non--hermitain Hamiltonian, see Ref. \cite{2020AIPA}.\\
If we set $\mathcal{O}=\mathcal{H}$, then Eq. (\ref{LVN}), or the commutator in Eq. (\ref{aver}), becomes the Liouville superoperator
\begin{equation}\label{Lrho}
    \frac{d \rho}{d t}=\mathscr{L}\rho=-i[\mathcal{H},\rho].
\end{equation}\label{Lcomponent}
Eq. (\ref{superH}) helps defining the components of $\mathscr{L}$ as
\begin{equation}
    \mathscr{L}_{\mu}^{~\nu}=-i\text{Tr}f^{\nu}[\mathcal{H},e_{\mu}]=-i\text{Tr}[f^{\nu},\mathcal{H}]e_{\mu}.
\end{equation}
Reintroducing the Liouville operator as a superoperator excavates its ``super power'' such that it manifestly plays the role of the Lie derivatives on the Riemannian $\rho$ manifold. The Jacobian of transformations between $\Sigma_{P}$ and $\Sigma_{Q}$ is given by
\begin{equation}
    J:=\text{det}\left[\frac{\partial\rho(t+\delta t)}{\partial\rho(t)}\right]=1-i\mathscr{L}\delta t
\end{equation}
or
\begin{equation}
    i\mathscr{L}_{\nu}^{~\mu}\rho^{\nu}:=\lim\limits_{\delta t\to 0}\frac{\rho^{\mu}(Q)(t+\delta t)-\rho^{\mu}_{0}(P)(t)}{\delta t}.
\end{equation}
It is obvious that, with the above geometric interpretation, we also can introduce the evolution superoperator
\begin{equation}\label{ULt}
    \mathscr{U}:= \exp[-i\mathscr{L}(\delta t)],
\end{equation}
which plays a role similar to that of the Lie groups, or their corresponding Killing fields, over the usual Riemannian manifolds.\\
We now introduce the Legendre transformation
\begin{eqnarray}
    \mathfrak{S}(\varrho_{\mu})&=&S-\langle\ln(\rho)\rangle \notag\\ 
    &=&S+\varrho_{\mu}\rho^{\mu}.
\end{eqnarray}
This transform reintroduces the $\rho$ components to be
\begin{equation}
    \rho^{\mu}:=\frac{\partial \mathfrak{S}}{\partial \varrho_{\mu}}.
\end{equation}
Therefore, the metric in Eq.(\ref{Qmetric}) can be contravariantized as
\begin{eqnarray}\label{contraQmetric}
    \mathpzc{G}^{\mu\nu}(\rho(\xi))&:=&\frac{\partial^2 \mathfrak{S}}{\partial \varrho_{\mu}\varrho_{\nu}} \notag\\
    &=&\frac{\partial\rho^{\kappa}}{\partial\varrho_{\mu}}\frac{\partial\rho^{\lambda}}{\partial\varrho_{\nu}}\frac{\partial^2 \mathfrak{S}}{\partial \rho^{\kappa}\partial\rho^{\lambda}}.
\end{eqnarray}
Consequently, Eq. (\ref{Qdistance}) is transformed into
\begin{equation}\label{d2S}
    d^{2}\mathfrak{S}=d^2S-\left[ d^2\varrho_{\mu}\rho_{\mu}+2d\varrho_{\mu}d\rho_{\mu}+\varrho_{\mu}d^2\rho_{\mu}\right].
\end{equation}
The metricity $\nabla\mathpzc{G}=0$, or the parallel transport along geodesics $\nabla_{\rho}\rho=0$, implies that there exists a set of connection coefficients $\{\}$ on the $\rho$ manifold similar to the Levi--Civita connections on the Einstein manifold. This means that both metric and connections are related through
\begin{subequations}
\begin{align}
     \mathpzc{G}(\nabla_{e^{\lambda}} e_{\mu},e_{\nu})&:=\big\{^{\;\kappa}_{\lambda\mu}\big\}\mathpzc{G}_{\kappa\nu}(\rho(\xi)), \label{GG}\\
     \big\{^{\;\kappa}_{\lambda\mu}\big\}=\frac{1}{2}\mathpzc{G}^{\kappa\iota}(\rho(\xi))\bigg[\partial_{\mu}\mathpzc{G}_{\iota\lambda}(\rho & (\xi))+ \partial_{\lambda}\mathpzc{G}_{\iota\mu}(\rho(\xi))-\partial_{\iota}\mathpzc{G}_{\mu\lambda}(\rho(\xi))\bigg], \label{GGG}
\end{align}
\end{subequations}
which is enough to introduce geodesic equations, Riemann curvature tensor and related other tensors. Additionally, Eq. (\ref{Qmetric}) and Eq. (\ref{contraQmetric}) reveal a Hessian structure on the density manifold such that Eq. (\ref{GG}) can be rearranged to get the corresponding connection coefficients of the first kind
\begin{align}\label{Connections}
    \big\{ {\scriptstyle \lambda\mu\nu} \big\}&=-\frac{1}{2}\frac{\partial^3 S}{\partial\rho^{\lambda}\partial\rho^{\mu}\partial\rho^{\nu}} \notag \\
    &=\frac{1}{2}\frac{\partial~ \mathpzc{G}_{\lambda\mu}(\rho(\xi))}{\partial\rho^{\nu}}.
\end{align}
In light of Balian \emph{et al.} metric, Eq. (\ref{Lrho}) and Eq. (\ref{ln.varrho}) yield
\begin{equation}\label{GLrho}
    \mathscr{L}^{~\nu}_{\mu}\varrho_{\nu}=\mathpzc{G}_{\mu\lambda}\mathscr{L}^{~\lambda}_{\nu}\rho^{\nu}.
\end{equation}
In light of Eq. (\ref{G}), the symmetric property of Balian \emph{et al.} metric, and the covariant form of the Liouville superoperator $\mathscr{L}_{\mu\nu}=\mathpzc{G}_{\mu\lambda}(\rho(\xi))\mathscr{L}^{\lambda}_{\nu}$, we differentiate Eq. (\ref{GLrho}) such that
\begin{equation}\label{partial.rho}
    0=\frac{1}{2}\frac{\partial~ \mathpzc{G}_{\mu\lambda}(\rho(\xi))}{\partial\rho^{\kappa}}\mathscr{L}^{~\lambda}_{\nu}\rho^{\nu}+\frac{1}{2}\left(\mathscr{L}_{\mu\kappa}+\mathscr{L}_{\kappa\mu}\right).
\end{equation}
Then, we substitute Eq. (\ref{Connections}) and Eq. (\ref{Lrho}) in Eq. (\ref{partial.rho}) such that
\begin{equation}
-i\frac{d\mathpzc{G}_{\mu\nu}(\rho(\xi))}{dt}=\mathscr{L}_{\mu\nu}+\mathscr{L}_{\nu\mu}~,
\end{equation}
which is the reason why we said before that $\mathscr{U}$, as defined in Eq. (\ref{ULt}), plays a role similar to that of Lie groups, or their corresponding Killing fields, over the usual Riemannian manifolds.

\subsection{The reduced spatial metric and the Euclidean structure of the space of observables}\label{Euclidstruct}

Since we deal with thermodynamical gauge degrees of freedom in information manifolds \cite{Souriau:1978}, it would be more convenient if we consider implementing Einstein's vierbein\footnote{Vierbeins in 4D manifolds become vielbeins in arbitrary dimensional manifolds.} (frame) field theory as a gauge field theory of gravity \cite{Einstein:1928}. We summarize this approach by defining an observable $\vec{\mathbb{W}}\equiv \mathbb{w}^{\mu}\mathpzc{e}_{\mu}$, where $\{\mathpzc{e}_{\mu}\}$ is the coordinate non--orthnormal bases set. Obviously $\mathpzc{G}_{\mu\nu}\equiv \mathpzc{e}_{\mu}\mathpzc{e}_{\nu}$. And for the non--coordinate orthnormal bases set $\{E_a\}$, as defined in the beginning of subsection (\ref{DensityManifold}), we have $E_{a}=e_a^{~\mu}\mathpzc{e}_{\mu}$, where $e_a^{~\mu}$ is the \textit{vielbein} structure
in such manifold. As expected $\eta_{ab}=\mathpzc{G}_{\mu\nu}~e_a^{~\mu}e_b^{~\nu}$. Therefore, we can connect this Euclidean space to the Riemannian density manifold through the spatial parts of \textit{vielbein} structure $e^{~\mu}_{i}$ such that the Euclidean flat metric corresponding to this structure is defined as $\delta_{ij}=\mathpzc{G}_{\mu\nu}~e_i^{~\mu}e_j^{~\nu}$, and $e_i^{~\mu}$ is indeed a function in the observable component $\mathbb{w}^i$. Moreover, the Euclidean metric $\delta_{ij}$ is the ($n-$1)--reduced dimensional metric of the $n$--dimensional Lorentzian metric $\eta_{ab}$. This strongly demands extracting the curved reduced spatial metric $\mathfrak{g}_{ij}$, that of a spatial slice, out of the metric $\mathpzc{G}_{\mu\nu}$. So for a vector $\vec{\mathbb{W}}$ that is tangent to the whole manifold and to a curve that lies in a slice $\Sigma_{(E_{a})}$ at constant orthnormal coordinate $E_{a}$, we define a \emph{normal} cotangent vector $n\equiv n_{\mu}E^{\mu}$, with components $n_{\mu}$, that is normal to $\Sigma_{(E_{a})}$ such that $n(\mathbb{W})=\mathbb{w}^{\nu}n_{\mu}=0$. And for constant temporal moment $t$ with $E_{0}$, the slices $\Sigma$'s become spatial, and the normal covector is defined as
\begin{subequations}\label{normalvector}
\begin{align}
n_{\mu}(t)&=\frac{E_0~e^{~0}_{\mu}}{\sqrt{\mathpzc{G}^{00}}}\\
\sigma\equiv\mathpzc{G}^{\mu\nu}n_{\mu}n_{\nu}&=n_{\mu}n^{\mu}=\pm 1
\end{align}
\end{subequations}
Now we define the projection operator to obtain the spatial metric $\mathfrak{g}_{ij}$. The projected part of $\mathbb{W}$ is defined as
\begin{equation*}
\forall~\mathbb{W}\in\mathpzc{T_\text{\emph{P}}(M)}, ~P\in\mathpzc{M},~\exists~\mathbb{W}_{{\perp}}\in T_P(\Sigma)\subset \mathpzc{T_\text{\emph{P}}(M)} ~\text{such that}~\mathbb{W}_{{\perp}}\equiv\mathscr{P}(\mathbb{W}).
\end{equation*}
This means $\{\mathpzc{e}_{\mu}\}$ and $\{E_a\}$ can be projected to $\{(\mathpzc{e}_{\perp})_{\mu}\}$ and $\{(E_{\perp})_a\}$ respectively as follows
\begin{subequations}
\begin{align}
&\mathbb{W}_{{\perp}}=[\mathscr{P}(\mathbb{W})]^{\mu}\mathpzc{e}_{\mu}=[\mathscr{P}^{~\mu}_{\nu}\mathpzc{e}_{\mu}]\mathbb{w}^{\nu}=\mathbb{w}^{\nu}(\mathpzc{e}_{{\perp}})_{\nu},\\
&\mathbb{W}_{\perp}=[\mathscr{P}(\mathbb{W})^{a}]E_{a}=[\mathscr{P}^{~a}_{b}E_{a}]\mathbb{w}^{b}=\mathbb{w}^{b}(E_{{\perp}})_{b}~.
\end{align}
\end{subequations}
Moreover, we can define another sets of bases $\{(\mathpzc{e}_{||})_{\mu}\}$ and $\{(E_{||})_a\}$ that are tangents to the slices $\Sigma_t$ as follows
\begin{subequations}\label{parallelbases}
\begin{align}
&(\mathpzc{e}_{||})_{\mu}=\mathpzc{e}_{\mu}-(\mathpzc{e}_{\perp})_{\mu}\\
&(E_{||})_{a}=E_{a}-(E_{\perp})_{a}\\
&\mathpzc{G}_{\mu\nu}=(\mathpzc{e}_{||})_{\mu}(\mathpzc{e}_{||})_{\nu}+(\mathpzc{e}_{\perp})_{\mu}(\mathpzc{e}_{\perp})_{\nu}
\end{align}
\end{subequations}
The density manifold has an Euclidean substructure as in the example mentioned at the end of subsection (\ref{vec.sp.construct}). This can help determining the relevant and the irrelevant part of the density \cite{Breuer:2007,Gemmer:2009}. Also,this means for any observable $\vec{\mathbb{W}}$, including the probability density,\footnote{Remember in the interaction picture the probability density can be treated as an observable operator.} every spatial relevant component $\mathbb{w}^{i}$ has orthogonal irrelevant projections $\mathscr{P}\vec{\mathbb{W}}$ such that
\begin{equation}\label{W-PW,W}
    \langle\vec{\mathbb{W}}-\mathscr{P}\vec{\mathbb{W}},\vec{\mathbb{W}}\rangle=\Big\langle\left(\vec{\mathbb{W}}-\mathfrak{o}\mathbb{w}\right)^{j},\mathbb{w}_{i}\Big\rangle=\delta^{~j}_{i}~,
\end{equation}
where the bilinear form is defined according to the map (\ref{G}), i.e. we can infer that $\mathscr{P}$ is a superoperator acting on both Euclidean and Riemannian manifolds. In order to obtain the components of this operator, we combine Eq. (\ref{normalvector}) and Eq. (\ref{parallelbases}) such that we get
\begin{equation}
    \mathscr{P}_{\mu}^{~\nu}=\delta_{\mu}^{~\nu}-\sigma n_{\mu}n^{\nu}.
\end{equation}
If we covariantize the projection operator, we get
\begin{equation}
    \mathscr{P}_{\mu\nu}=\mathpzc{G}_{\mu\nu}-\sigma n_{\mu}n_{\nu}.
\end{equation}
This helps defining the metric in form of the \emph{covariant} projection operator and the normal vectors as
\begin{equation}
    \mathpzc{G}_{\mu\nu}=\mathscr{P}_{\mu\nu}+\sigma n_{\mu}n_{\nu}.
\end{equation}
This means that the covariant projection operator defines the reduced spatial metric $\mathfrak{g}_{ij}$ if we realize that $\mathscr{P}$ is made only of spatial components $\mathscr{P}_{ij}$. So we apply $n^in^j$ on both sides of the last equation, with the help of the normal vector definition in Eq. (\ref{normalvector}), to get 
\begin{subequations}
    \begin{align}
        \mathfrak{g}_{ij}\equiv e^{~\mu}_{0}e^{~0}_{i}&~\mathpzc{G}_{\mu\nu}~e^{~\nu}_{0}e^{~0}_{j}=\mathpzc{G}_{\mu\nu}\delta^{~\mu}_{i}\delta^{~\nu}_{j}, \label{spatialFisher}\\        &\mathfrak{g}^{jk}\mathfrak{g}_{ki}=\delta^{~j}_{i}.
    \end{align}
\end{subequations}
In an information manifold this metric plays the same role the spatial spacetime metric $\gamma_{ij}$ does in defining the expansion parameter in Eq. (\ref{theta.gamma}). It is worth noting that the vielbein acts on the density vector as a projector operator to yield the components of the density vector in the Liouville space, which is another way to define the axes in Fig. \hyperref[fig1]{(1)}, i.e. we could start from the projector operator and the vielbein structure backward until we reach the Liouville vector representation of the density operator; both approaches therefore are equivalent. So if we return back to Fig. \hyperref[fig1]{(1)} and choose a point $P'(t_0)\in\Sigma_Q$ along the curve $t_0$, then
\begin{subequations}\label{projection.rho}
\begin{align}
    \mathscr{P}\rho(t_0)&=\rho_0(t_0),\\
    \rho^i:=\langle \mathbb{w}^i\rangle&=\langle \mathbb{w}^i,\rho\rangle=\langle \mathbb{w}^i,\rho_0\rangle,~~ \forall i.
\end{align}
\end{subequations}
Now it is safe to infer that the distance on the surfaces $\Sigma$ in Fig. \hyperref[fig1]{(1)} are given by
\begin{eqnarray}
ds^2_{\Sigma}&=&\mathfrak{g}_{ij}(\rho(\xi))d\rho^id\rho^j \notag\\
&=&\mathfrak{g}^{ij}(\rho(\xi))d\rho_id\rho_j~,
\end{eqnarray}
 where $\mathfrak{g}_{ij}(\rho(\xi))$ could be not equal to $\mathfrak{g}^{ij}(\rho(\xi))$ in general, i.e. $d\rho_i\neq\mathfrak{g}_{ij}(\rho(\xi))d\rho^{j}$ necessarily as $d\rho_i$ is more like the component $\mathfrak{o}_i$ of $\vec{\mathcal{O}}$ as defined previously. Another important manifold structure can be obtained by combining Eq. (\ref{tangent.rho}) and Eq.(\ref{projection.rho}) such that
\begin{equation}
    \langle \mathbb{w}^i,\scripty{r} \rangle=0~,
\end{equation}
where the vector $\scripty{r}=(\rho-\rho^i_0)$ is the tangent along the curve $t_0$, see Fig. \hyperref[fig1]{(1)}. Thus, we have a \textit{vector bundle} structure, where the \textit{base} is the surface $\Sigma_P$ and the \textit{fibres} are the curves $t_0+n\delta t, n\in \mathbb{N}$. See Ref. \cite{Guo:2021fax} for more about the relation with the \emph{blurred space}.

\subsection{Fisher metric and Kullback-Leibler divergence}\label{FisherKLdiv}

The previously constructed density manifold has an Euclidean signature. Then, we target constructing an information manifold with a \textit{Lorentzian signature} diag$(-1,+1,+1,\cdots)$, and we achieve this goal in subsection (\ref{pseudomanifold}). For now, we focus on relating Balian \emph{et al.} metric to Fisher metric. As the vacuum spacetime is in a ``continuous'' experience of quantum fluctuations, it can be optimized stochastically such that the expectation values of the operators over spacetime, i.e. the stochastic variables,\footnote{For more on the properties of the stochastic variables, see Ref. \cite{Namsrai:1986md}.} are invariant under coordinate transformations between different frames of references \cite{Lindgren:2019tdd}. This means we can define the density vectors $\rho^{\mu}$, which is a function in the classical variables $x^{\mu}$ characterizing the spacetime itself, as function in the stochastic variables $\mathbb{X}^{\mu}\equiv \mathbb{X}^{\mu}(\langle x^{\mu}\rangle,\sigma_{x^{\mu}})$ that are functions in a spatial space of averages $\mathbb{X}^{\mu}(\langle x^{\mu}\rangle)$ and standard deviations $\sigma_{x^{\mu}}$.\footnote{See Ref. \cite{Amari:1994} for more on the $2D$ manifold of averages and standard deviations in the exponential families of normal distribution.} Then, we may guess that Balian \emph{et al.} metric $\mathpzc{G}_{\mu\nu}\left(\rho(\xi^{\mu})\right)$ in Eq. (\ref{Qdistance}) could be expressed as an explicit function of $\xi^{\mu}$ and $\mathbb{X}^{\mu}$ variables, i.e. $\mathpzc{G}_{\mu\nu}\equiv\mathpzc{G}_{\mu\nu}(\xi^{\mu};\mathbb{X}^{\mu})$. In order to check the validity of this guess, we need to check that this Ruppeiner metric maintains the same form when coordinate transformations are considered. The best candidate to test this requirement is the Kullback--Leibler divergence
\begin{equation}\label{KLdiv}
    D_{\text{KL}}=\sum\limits_{\xi^{i};\mathbb{X}^{i}}\rho(\xi^{\mu};\mathbb{X}^{\mu})\ln\frac{\rho(\xi^{\mu};\mathbb{X}^{\mu})}{\rho(\xi^{\mu}+d\xi^{\mu};\mathbb{X}^{\mu})}\Delta\mathbb{X}^{i}.
\end{equation}
As the spacetime variables are infinitesimally changing and $\rho\to d\rho$, then Eq. (\ref{KLdiv}) becomes
\begin{align}\label{IntKLdiv}
    D_{\text{KL}}=-\int\limits_{\mathcal{M}} & \rho(\xi;\mathbb{X})\Big[ \frac{1}{\rho(\xi;\mathbb{X})}\frac{\partial \rho(\xi;\mathbb{X})}{d\xi^{\mu}}d\xi^{\mu}
    -\frac{1}{2}\frac{1}{\rho^2(\xi;\mathbb{X})}\frac{\partial \rho(\xi;\mathbb{X})}{d\xi^{\mu}}\frac{\partial \rho(\xi;\mathbb{X})}{d\xi^{\nu}}d\xi^{\mu}d\xi^{\nu}\Big]d\mathbb{X},
\end{align}
where $\rho(\xi;\mathbb{X})\equiv\rho(\xi^{\mu};\mathbb{X}^{\mu})$. The second term in Eq. (\ref{IntKLdiv}) is nothing but the \textit{Fisher metric} 
\begin{equation}
    ds^2:=\sum\limits_i\frac{(d\rho_i)^2}{\rho_i}~,
\end{equation}
which is the same metric in Eq. (\ref{Qdistance}) but as an explicit function in $\xi^{\mu}$ and $\mathbb{X}^{\mu}$ \cite{Balian:1986jrj}. Applying Fisher metric, chain rule, and the completeness relation of the density to Eq. (\ref{KLdiv}) yields
\begin{equation}\label{KLmetric}
    D_{\text{KL}}=\int\limits_{\mathcal{M}}\frac{1}{2}\mathpzc{G}_{\mu\nu}(\xi;\mathbb{X})d\xi^{\mu}d\xi^{\nu}d\mathbb{X}.
\end{equation}
Eq. (\ref{KLdiv}) can be seen as
\begin{align}\label{DKLd2S}
    D_{\text{KL}}\sim -d^2S=d\bigg[\frac{\partial S}{\partial \rho_{\mu}}d\rho^{\mu}\bigg]=\frac{\partial^2 S}{\partial \rho^{\mu}\partial \rho^{\nu}}\frac{\partial \rho^{\mu}}{\partial \xi^{\kappa}}\frac{\partial \rho^{\nu}}{\partial \xi^{\lambda}}d\xi^{\kappa}d\xi^{\lambda}.
\end{align}
By comparing Eq. (\ref{contraQmetric}), Eq. (\ref{KLmetric}), and Eq. (\ref{DKLd2S}), and as the KL divergence is nothing but a modified Shannon entropy, the covariant form of Eq. (\ref{contraQmetric}) says that
\begin{equation}\label{ShanonKL}
   \mathpzc{G}_{\mu\nu}(\varrho(\xi;\mathbb{X}))=\frac{\partial \varrho}{\partial \xi^{\mu}}\frac{\partial \varrho}{\partial \xi^{\nu}}.
\end{equation}
As we notice, the density is no longer a vector, it is just a function, and the vectors of the new manifold are $\partial_{\mu}\equiv\partial/\partial\xi^{\mu}$. If we strict expressing all densities as as functions in stochastic variables $\mathbb{X}^{\mu}$ rather than the canonical ones $\xi^{\mu}$, i.e. $\rho\equiv\rho(\mathbb{X}^{\mu})$, then, with help of Eq. (\ref{aver}--\ref{bilinear}), we suppress Fisher metric density $\xi^{\mu}$--dependence, i.e. from now on the Balian \emph{et al.} metric\footnote{We move from the information manifold of Balian \emph{et al.} \cite{Balian:1986jrj} to the information manifold of Amari \cite{Amari:2000}.} $\mathpzc{G}_{\mu\nu}$ is not the metric we use. And Eq. (\ref{ShanonKL}) should be improved as
\begin{equation}\label{Fisher}
    \langle\mathpzc{G}_{\mu\nu}(\xi;\mathbb{X})\rangle=\langle\frac{\partial \varrho}{\partial \xi^{\mu}} \frac{\partial \varrho}{\partial \xi^{\nu}}\rangle.
\end{equation}
In order to understand the last result, we need to get back to the Fisher metric. Without loss of generality, Eq. (\ref{IntKLdiv}) describes the relation between the entropy at the reduced state $\rho_0$ and any other state $\rho$, see the Points $P$ and $Q$ in Fig. \hyperref[fig1]{(1)}. Then, \textit{in information manifold the probabilistic average of the stochastic Fisher metric $\mathbb{g}_{\mu\nu}$ plays the same role the spacetime metric does in the Riemannian manifolds}, i.e. $g_{\mu\nu}\simeq \mathbb{g}_{\mu\nu}$. Thus,
\begin{equation}\label{FisherShanon}
    \mathbb{g}_{\mu\nu}:=\langle\mathpzc{G}_{\mu\nu}(\xi,\mathbb{X})\rangle~.
\end{equation}
So we confirm that the metric of the information manifold maintains its form when coordinate transformations are considered. This is crucial for defining Einstein tensor and the analogue of the gravitational constant in the information manifold as we will see by the end of the next subsection (\ref{HessianEinsteinTensor}).

\subsection{Hessian structure and Einstein tensor}\label{HessianEinsteinTensor}

For an information manifold endowed with exponential family of distribution, there exists a potential function $\phi$ such that its Hessian defines the metric of that manifold \cite{Ciaglia}. Non--exponential families do the same but in a less straightforward way,\footnote{This is mainly because the Bayesian hypothesis \cite{e22101100} related to the Hessian in non--exponential families needs variational Bayesian approximations, like Laplace approximation and parameter separation parameterization, to obtain the posterior distribution \cite{Hui2014}.} for more on that see Ref. \cite{6002511, MATSUZOE} and references there. For simplicity we discuss the Hessian structure of exponential families, but our discussion is applicable to non--exponential ones too. We share with Ref. \cite{Matsueda:2013saa} constructing Einstein tensor from the Fisher metric in information manifold. Also, we find Einstein tensor to be endowed with relevant information to construct the energy--momentum tensor from the varying cumulant partition function defined as a scalar field. This is guaranteed naturally since the entropy data of the system underpin the field strength of the system. Thus, we can consider Einstein equations in information manifold as the equations of coarse--grained states for the original microscopic system of quantum field theories behind the classical ones.
The major difference between this work and the endeavor of Ref. \cite{Matsueda:2013saa} is that, without assuming any family of exponentials, we find a \emph{positively definite cosmological--like background term} in the coarse--grained Einstein equations, particularly in the terms containing the derivatives of the Christoffel connections. This suggests redefining Einstein tensor in information manifold to become a Lovelock tensor. Consequently, coarse--graining the states could reveal extra disguised higher ordered curvature terms in the theory.
Additionally, in contrary to Ref. \cite{Matsueda:2013saa} choice of real probability distributions that suit AdS/CFT, we impose the family of exponentials to be defined as complex probability distributions such that we construct Einstein tensor in a information manifold with Lorentzian signature. This does not change the result that the gravitational constant is indeed dynamical. However, \emph{it changes the sign of the extremal area and the entropy}. This result is consistent with the fact that complex probabilities are associated with non--Hermitian Hamiltonians and their non--unitary transformations.
Ref. \cite{Matsueda:2013saa} admits the problem of controlling the energy scale of the quantum field theory in the information geometric approach. In order to resolve that, we suggest repeating the process coarse--graining until fine--graining is achieved \cite{Safranek}. Consequently, the information approach would have a renormalization process, and the von Neumann equation stays fine--grained as suggested in Ref. \cite{Maldacena:2020ady}. Thus, our modifications may introduce quantum information geometry approach to dS/CFT \cite{Strominger:2001pn,Bousso:2002ju} with the help of the obtained pseudo--entropy \cite{Doi:2022iyj,Doi:2023zaf}.

As shown in Appendix 2.7 of Ref. \cite{Hui2014}, the variational Bayesian inferences limited to exponential families, like the KL divergence, can be generalized through the technique of \emph{parameter
separation parameterization}. The technique is applicable to both exponential and non--exponential families of distributions by \emph{linearly} relating $\rho$ to a \emph{real valued potential function}.\footnote{See function $h(y)$ defined in Eq. 2.14 in Ref. \cite{Hui2014}.} The potential function $\phi$ would help defining the metric of the information manifold as shown in the \hyperref[append]{Appendix}. Without loss of generality, the equality $\partial_{\mu}\partial_{\nu}\phi=\mathbb{g}_{\mu\nu}$ in Eq. (\ref{Hessian}) is very similar to the bilinear relation $\text{Hess}(\phi)=\nabla\nabla\phi$, or the \textit{Hessian}, in Riemannian geometry \cite{Amari:2000}. We are allowed to do this comparison because of the bundle structure in the density manifold we referred to at the end of subsection (\ref{Euclidstruct}). Remember that $\varrho\in\text{T}^{*}_P\Sigma$, i.e. the Hessian acts on the potential function $\phi$ to get an element in the sections of tangent bundles $\text{Hess}(\phi)\in\Gamma(\text{T}^{*}\Sigma\otimes \text{T}^{*}\Sigma)$. Also, the density manifold and the canonical variable manifold, which could be the conventional spacetime, are related through transforming $\varrho$ into $\partial_{\mu}\varrho$. Then, we can apply a derivative on Eq. (\ref{Hessianrho}) to get
\begin{align}\label{partvarrho3}
    \partial_{\lambda}\mathbb{g}_{\mu\nu}=\partial_{\lambda}\partial_{\mu}\partial_{\nu}\phi
    =\langle\partial_{\lambda}(\partial_{\mu}\partial_{\nu}\varrho)\rangle=\langle\partial_{\lambda}(\partial_{\mu}\varrho\partial_{\nu}\varrho)\rangle=-\langle\partial_{\lambda}\varrho\partial_{\mu}\varrho\partial_{\nu}\varrho\rangle,
\end{align}
where the tensor relation between the vectors $\partial_{\mu}\varrho$ is suppressed, the last equality comes from $\langle\partial_{\mu}(\partial_{\nu}\varrho)\rangle=\langle\partial_{\mu}\varrho\partial_{\nu}\varrho\rangle$, and  the negative sign in last equality comes from the metric definition in Eq. (\ref{Sbilinear}). More obviously from the behavior of the derivatives, we have
\begin{equation}
   \partial_{\lambda}\mathbb{g}_{\mu\nu}=\langle\partial_{\lambda}\partial_{\mu}\varrho\partial_{\nu}\varrho\rangle+\langle\partial_{\mu}\varrho\partial_{\lambda}\partial_{\nu}\varrho\rangle.
\end{equation}
Using the symmetric property of $\mathbb{g}_{\mu\nu}$, the last two equations lead us to
\begin{equation}
    \partial_{\lambda}\mathbb{g}_{\mu\nu}=\frac{1}{2}\Big[\langle\partial_{\lambda}\partial_{\mu}\varrho\partial_{\nu}\varrho\rangle+\langle\partial_{\mu}\varrho\partial_{\lambda}\partial_{\nu}\varrho\rangle-\langle\partial_{\lambda}\varrho\partial_{\mu}\varrho\partial_{\nu}\varrho\rangle\Big].
\end{equation}
In light of Eq. (\ref{GGG}), the last result is very enticing to define the Christoffel connection corresponding to the canonical variable manifold as \cite{Amari:1994,Amari:2000}
\begin{align}\label{Christ}
    \mathbb{\Gamma}^{\lambda}_{\mu\nu}&=-\frac{1}{2}\mathbb{g}^{\lambda\kappa}\partial_{\kappa}\partial_{\mu}\partial_{\nu}\phi \notag\\
    &=\mathbb{g}^{\lambda\kappa}\Big[\langle\partial_{\kappa}\varrho\partial_{\mu}\partial_{\nu}\varrho\rangle-\frac{1}{2}\langle\partial_{\kappa}\varrho\partial_{\mu}\varrho\partial_{\nu}\varrho\rangle\Big].
\end{align}
Next, we use Eq. (\ref{Christ}) to calculate the Ricci tensor and the Ricci scalar as functions in the dual density variables and their derivatives. The Ricci tensor is obtained as usual from the contracted Riemann tensor
\begin{equation}\label{RicTen}
    \mathbb{R}_{\mu\nu}= \mathbb{g}^{\alpha\beta}\mathbb{R}_{\alpha\mu\beta\nu}=\partial_{\alpha}\mathbb{\Gamma}^{\alpha}_{\mu\nu}-\partial_{\nu}\mathbb{\Gamma}^{\alpha}_{\mu\alpha}+\mathbb{\Gamma}^{\alpha}_{\beta\alpha}\mathbb{\Gamma}^{\beta}_{\mu\nu}-\mathbb{\Gamma}^{\alpha}_{\beta\nu}\mathbb{\Gamma}^{\beta}_{\mu\alpha}.
\end{equation}
Meanwhile the Ricci scalar also is obtained as usual for the contacted Ricci tensor
\begin{equation}
    \mathbb{R}= \mathbb{g}^{\mu\nu}\mathbb{R}_{\mu\nu}.
\end{equation}
Then, the Einstein tensor in information manifold becomes
\begin{equation}\label{EinEqInfoGeometry}
    \mathbb{G}_{\mu\nu}=\mathbb{R}_{\mu\nu}-\frac{1}{2}\mathbb{g}_{\mu\nu}\mathbb{R}.
\end{equation}
As shown in the \hyperref[append]{Appendix}, we deconstruct Eq. (\ref{EinEqInfoGeometry}) into two main pieces. We analyze these pieces using the definitions of the metric and Christoffel connections in information manifolds such that the Hessian structure is realized. So, we can reintroduce Eq. (\ref{EinEqInfoGeometry}) as
\begin{subequations}
\begin{align}
\mathbb{G}_{\mu\nu}=\Lambda\mathbb{g}_{\mu\nu}+(\tilde{\mathbb{R}}_{\mu\nu}&-\frac{1}{2}\mathbb{g}_{\mu\nu}\tilde{\mathbb{R}})+\frac{1}{2}\mathbb{g}_{\mu\nu}\tilde{\mathbb{R}},\\
\text{or}~~\tilde{\mathbb{G}}_{\mu\nu}+\Lambda\mathbb{g}_{\mu\nu}&=\mathbb{G}_{\mu\nu}-\frac{1}{2}\mathbb{g}_{\mu\nu}\tilde{\mathbb{R}},
\end{align}
\end{subequations}
where $\tilde{\mathbb{R}}_{\mu\nu}$ is explained in Eq. \ref{mathcalR}, and $\tilde{\mathbb{G}}_{\mu\nu}+\Lambda\mathbb{g}_{\mu\nu}$ plays a rule similar to that of Lovelock tensor $A_{\mu\nu}=G_{\mu\nu}+\Lambda g_{\mu\nu}$ in Einstein spacetime manifold \cite{Lovelock:1972vz}. 

The first new piece obtained from rearranging then deconstructing Eq. (\ref{EinEqInfoGeometry}), which is Eq. (\ref{CosmInfoD}-\ref{Lambda(D/2-1)append}), gives a cosmological-like term in the information manifold as
\begin{equation}\label{Lambda(D/2-1)}
    \Big(\partial_{\alpha}\mathbb{\Gamma}^{\alpha}_{\mu\nu}-\partial_{\nu}\mathbb{\Gamma}^{\alpha}_{\mu\alpha}\Big)-\frac{1}{2}\mathbb{g}_{\mu\nu}\mathbb{g}^{\kappa\lambda}\Big(\partial_{\alpha}\mathbb{\Gamma}^{\alpha}_{\kappa\lambda}-\partial_{\kappa}\mathbb{\Gamma}^{\alpha}_{\lambda\alpha}\Big)=\frac{1}{2}D(\frac{D}{2}-1)\mathbb{g}_{\mu\nu}\equiv\frac{1}{2}\Lambda \mathbb{g}_{\mu\nu}~,
\end{equation}
where $\Lambda=D(D/2-1)$ is defined for $D\geqslant 2$ dimensional manifold as in Lovelock theory of gravity \cite{Lovelock:1971yv}. This helps our information manifold pass the necessary condition, but not sufficient on its own, to develop an additional Gauss--Bonnet term for the corresponding Riemann tensor that is expressed in its information form. The appearance of $D(D/2-1)$ term in the theory is very tempting to study the effects of having higher--curvature terms in the context of holography. We leave that to be discussed hopefully in a future study.
We notice that the cosmological constant we obtained does not depend on the manifold parameters. It only depends on the number of dimensions, and it vanishes when we consider $D=2$. This may point to background symmetry behind the cosmological constant, and it may have a relation with vanishing cosmological constant in the context of some M/string theory \cite{Witten:2000zk}. More importantly, this piece in the $\Tilde{\mathbb{G}}_{\mu\nu}$ tensor should exist for all kinds of exponential or non--exponential probability distributions. It neither demands $\rho$ to be exponential, Gaussian or complex one, nor to be in dS or AdS spaces. And this is our major difference between this work and the endeavors in Ref. \cite{Matsueda:2013saa}.

The second piece after deconstructing Eq. (\ref{EinEqInfoGeometry}) appears written in Eq. (\ref{mathcalRappend}) in the \hyperref[append]{Appendix}. This piece of non differentiated Christoffels, obtained from both $\mathbb{R}_{\mu\nu}$ and $\mathbb{g}_{\mu\nu}\mathbb{R}$, defines a new tensor
\begin{align}\label{mathcalR}
    \mathbb{\widetilde{R}}_{\mu\nu}&=
    \Big(\mathbb{\Gamma}^{\alpha}_{\beta\alpha}\mathbb{\Gamma}^{\beta}_{\mu\nu}-\mathbb{\Gamma}^{\alpha}_{\beta\nu}\mathbb{\Gamma}^{\beta}_{\mu\alpha}\Big)-\frac{1}{2}\mathbb{g}_{\mu\nu}\mathbb{g}^{\kappa\lambda}\Big(\mathbb{\Gamma}^{\alpha}_{\beta\alpha}\mathbb{\Gamma}^{\beta}_{\kappa\lambda}-\mathbb{\Gamma}^{\alpha}_{\beta\kappa}\mathbb{\Gamma}^{\beta}_{\lambda\alpha}\Big) \notag\\
    &=\frac{1}{4}\mathbb{g}^{\alpha\kappa}\mathbb{g}^{\beta\lambda}\Big[\langle\partial_{\alpha}\varrho\partial_{\beta}\varrho\partial_{\kappa}\varrho\rangle\langle\partial_{\mu}\varrho\partial_{\nu}\varrho\partial_{\lambda}\varrho\rangle-\langle\partial_{\nu}\varrho\partial_{\beta}\varrho\partial_{\kappa}\varrho\rangle\langle\partial_{\mu}\varrho\partial_{\alpha}\varrho\partial_{\lambda}\varrho\rangle\Big].
\end{align}
Without loss of generality,\footnote{Generality is not lost as the crucial steps in Eq.(\ref{Hessian}--\ref{Hessianrho}) are guaranteed by applying previously mentioned technique of parameter separation parametrization on non--exponential families \cite{Hui2014}.} we follow the exponential family example in Ref. \cite{Matsueda:2013saa}. We use Eq. (\ref{ExpBeta}) and after so Eq. (\ref{mathcalR}) becomes
\begin{align}\label{widetildeR}
    \mathbb{\widetilde{R}}_{\mu\nu}=\frac{1}{4}\mathbb{g}^{\alpha\kappa}\mathbb{g}^{\beta\lambda}\Bigg\{&\Big[\mathpzc{G}_{\alpha\beta}\mathpzc{G}_{\kappa\lambda}-\mathbb{g}_{\alpha\beta}\mathbb{g}_{\kappa\lambda}\Big] 
     +\mathpzc{G}_{\kappa\lambda}\Big[\dot{\mathpzc{G}}_{\alpha\beta}\langle\mathbb{X}-\langle x\rangle\rangle+\frac{1}{2}\ddot{\mathpzc{G}}_{\alpha\beta}\langle(\mathbb{X}-\langle x\rangle)^2\rangle+\cdots\Big] \notag\\
    & +\mathpzc{G}_{\alpha\beta}\Big[\dot{\mathpzc{G}}_{\kappa\lambda}\langle\mathbb{X}-\langle x\rangle\rangle+\frac{1}{2}\ddot{\mathpzc{G}}_{\kappa\lambda}\langle(\mathbb{X}-\langle x\rangle)^2\rangle+\cdots\Big]\Bigg\}\times\partial_{\mu}\phi\partial_{\nu}\phi.
\end{align}
Or
\begin{equation}
    \mathbb{\widetilde{R}}_{\mu\nu}=\frac{1}{2}\mathbb{G}_D~\partial_{\mu}\phi\partial_{\nu}\phi.
\end{equation}
where the dynamical entity $\mathbb{G}_D$ is defined as
\begin{align}
    \mathbb{G}_D=\frac{1}{2}\mathbb{g}^{\alpha\kappa}\mathbb{g}^{\beta\lambda}\Bigg\{&\Big[\mathpzc{G}_{\alpha\beta}\mathpzc{G}_{\kappa\lambda}-\mathbb{g}_{\alpha\beta}\mathbb{g}_{\kappa\lambda}\Big] 
     +\mathpzc{G}_{\kappa\lambda}\Big[\dot{\mathpzc{G}}_{\alpha\beta}\langle\mathbb{X}-\langle x\rangle\rangle+\frac{1}{2}\ddot{\mathpzc{G}}_{\alpha\beta}\langle(\mathbb{X}-\langle x\rangle)^2\rangle+\cdots\Big] \notag\\
    & +\mathpzc{G}_{\alpha\beta}\Big[\dot{\mathpzc{G}}_{\kappa\lambda}\langle\mathbb{X}-\langle x\rangle\rangle+\frac{1}{2}\ddot{\mathpzc{G}}_{\kappa\lambda}\langle(\mathbb{X}-\langle x\rangle)^2\rangle+\cdots\Big]\Bigg\},
\end{align}
where
\begin{equation}
    \dot{\mathpzc{G}}_{\mu\nu}(\langle x\rangle)=\lim\limits_{\mathbb{X}\to\langle x\rangle}\frac{\partial}{\partial \mathbb{X}}{\mathpzc{G}}_{\mu\nu}(\mathbb{X})
\end{equation}
and $\ddot{\mathpzc{G}}_{\mu\nu}(\langle x\rangle)$ is the usual second derivative of the Taylor expansion expressed in Eq. (\ref{expandedG}). The limit closes the stochastic variable to the more probable value of the random variable.\\
Using the cumulant partition function $\phi$ as a scalar field in the information manifold, we can define a Lagrangian in $D=4$ for an effective field theory as
\begin{equation}\label{effectiveL}
    \mathbb{L}=\mathbb{g}^{\mu\nu}\partial_{\mu}\phi\partial_{\nu}\phi,
\end{equation}
and the corresponding energy--momentum tensor is
\begin{equation}\label{EinsteinInfo}
    \mathbb{T}_{\mu\nu}=\mathbb{g}_{\mu\lambda}\frac{\partial \mathbb{L}}{\partial(\partial_{\lambda}\phi)}\partial_{\nu}\phi-\mathbb{g}_{\mu\nu}\mathbb{L}=\frac{1}{\mathbb{G}_4}\Bigg(\mathbb{\widetilde{R}}_{\mu\nu}-\frac{1}{2}\mathbb{g}_{\mu\nu}\mathbb{\widetilde{R}}\Bigg)=\frac{1}{\mathbb{G}_4}\mathbb{\widetilde{G}}_{\mu\nu}~,
\end{equation}
where obviously the last equation is Einstein equation with its \textit{reduced} tensors $\mathbb{\widetilde{G}}_{\mu\nu}$ and $\mathbb{\widetilde{R}}_{\mu\nu}$ as functions in the stochastic variables $\mathbb{X}$ in such information geometry. Also, we can add the cosmological constant term we obtained before in Eq. (\ref{Lambda(D/2-1)}). Comparing Eq. (\ref{EinsteinInfo}) with Eq. (\ref{Einstein}), we get
\begin{equation}\label{G4infomanifold}
\mathbb{G}_4\simeq\frac{8\pi\ell^2_P}{\hbar},
\end{equation}
which again is nothing but the gravitational constant in the information form for such $4D$ geometry. This indicates quantum information geometry induces gravity phenomena, and the gravitational constant is no longer constant besides its $\hbar$ dependence. This may support studies that implies inducing gravity from quantum mechanics \cite{VanRaamsdonk:2010pw,Steinacker:2010rh,Padmanabhan:2007tm,Beggs:2013pxa,Das:2021ope}. Also, It could be linked to realizing space and time as approximate macroscopic concepts stemming fundamentally from quantum field theories \cite{Seiberg:2006wf}. Additionally, a varying gravitational constant may give a clue to Dirac's large numbers hypothesis \cite{Dirac:1978xh} that also implies a varying gravitational constant based on the simple analysis of dimensionless constants that are provided by nature. 

The results obtained in Eq. (\ref{Christ}--\ref{Lambda(D/2-1)}) are discussed in details in the \hyperref[append]{Appendix} of the manuscript. There, the family of exponentials is not assumed to be either complex, which is what we discuss in the next subsection, or real as in Ref. \cite{Matsueda:2013saa}, since we keep everything in the appendix in terms of an arbitrary $\varrho$. Additionally in Ref. \cite{Matsueda:2013saa}, after the Gaussian exponential was imposed, the ``whole'' Einstein tensor is equated to a ``negative'' cosmological constant. This is not what we obtain. Rather, we say that the Einstein tensor $\mathbb{G}_{\mu\nu}$ should be split into $\partial \mathbb{\Gamma}$ terms that correspond to a ``positively'' cosmological constant defined in terms of the dimension $D$ as in Eq. (\ref{Lambda(D/2-1)}), and $\mathbb{\Gamma}\mathbb{\Gamma}$ terms that introduce a modified Ricci tensor $\Tilde{ \mathbb{R}}_{\mu\nu}$ as the kinetic term $\partial_{\mu}\phi\partial_{\nu}\phi$ in Eq. (\ref{widetildeR}). Notice that the Lagrangian $\mathbb{L}\sim\mathbb{g}^{\mu\nu}\Tilde{\mathbb{R}}_{\mu\nu}\sim\Box\phi$ renders the reduced Einstein tensor $\Tilde{\mathbb{G}}_{\mu\nu}$ upon applying the variational principle with respect to $\phi$, while the $\Lambda\mathbb{g}_{\mu\nu}$ piece is produced from varying $\mathbb{L}$ with respect to $\mathbb{g}_{\mu\nu}=\partial_{\mu}\partial_{\nu}\phi$. This reminds us with the $Y$ piece in the modified Lagrangian of gravity in the framework of superstrings \cite{Gross:1986iv}, both share the same Hessian structure. In brief, we suggest that the Einstein tensor $G_{\mu\nu}$ in Ref. \cite{Matsueda:2013saa} should rather be treated as a Lovelock tensor $A_{\mu\nu}=G_{\mu\nu}+\Lambda g_{\mu\nu}$. Such tensor should be split into a cosmological term $\Lambda\mathbb{g}_{\mu\nu}$ and a modified Ricci tensor $\Tilde{\mathbb{R}}_{\mu\nu}$, and the later can be used to introduce a new modified Einstein tensor $\Tilde{\mathbb{G}}_{\mu\nu}$ that has no cosmological terms in it. All that can be obtained without assuming the densities to be expressed as family of exponentials or any other family.

\subsection{Obtaining a pseudo-Riemannian information manifold}\label{pseudomanifold}

A question remains about how to construct an arbitrary $(D+1)$--dimensional information geometry with a \textit{Lorentzian signature} diag$(-1, +1, \cdots)$. And without loss of generality, we follow the complex Gaussian ansatz in Ref. \cite{Calmet:2004a}, and we comment on the consequences of this ansatz in the \hyperref[conclusion]{discussion} section, to get a $(1+1)$--dimensional information manifold with a Lorentzian signature diag$(-1, +1)$ by assuming the following distribution
\begin{equation}\label{Exp.Complex.t}
    \rho (t)=\frac{1}{\sqrt{2\pi}\sigma}\exp\left[ -\frac{(t-i\langle t\rangle)^2}{2\sigma}\right]=\exp\left[-\ln(\sqrt{2\pi}\sigma)-\frac{t^2}{2\sigma^2}+\frac{it\langle t\rangle}{\sigma^2}+\frac{\langle t\rangle^2}{2\sigma^2}\right],
\end{equation}
which requires components of the corresponding Fisher metric to be
\begin{subequations}
\begin{align}
    \mathbb{g}_{00}&=\int d^2\xi \frac{1}{\rho(\xi)}\left(\frac{\partial\rho(\xi)}{\partial\xi^0} \right)^2=-1,\\
    \mathbb{g}_{11}&=\int d^2\xi \frac{1}{\rho(\xi)}\left(\frac{\partial\rho(\xi)}{\partial\xi^1} \right)^2=1,\\
    \mathbb{g}_{01}&=\int d^2\xi \frac{1}{\rho(\xi)}\left(\frac{\partial\rho(\xi)}{\partial\xi^0} \right)\left(\frac{\partial\rho(\xi)}{\partial\xi^1} \right)=0.
\end{align}
\end{subequations}
The Lorentzian signature of the metric requires defining a \textit{complex exponential family} as in Eq. (\ref{Exp.Complex.t}) and maintains the general form
\begin{equation}\label{ExpFam}
    \rho(\xi^{\mu})=\exp\left[\xi^{\mu} \mathbb{E}_{\mu}(\mathbb{X}^{\nu})-\phi(\xi^{\nu})\right],
\end{equation}
where the function $\mathbb{E}_{\mu}(\mathbb{X}^{\nu})$ plays the role of any physical property related to the corresponding canonical or intensive variables $\xi^{\nu}$, and $\phi(\xi^{\nu})$ is the \textit{cumulant partition function} of the $\mathbb{E}_{\mu}(\mathbb{X}^{\nu})$ states \cite{Amari:2000}, see Eq. (\ref{ExpBeta}) and after in the \hyperref[append]{Appendix}. By comparing Eq. (\ref{Exp.Complex.t}) to Eq. (\ref{ExpFam}), it is easy to notice that
\begin{subequations}
\begin{align}
    \mathbb{E}_{\mu}&:=(\mathbb{E}_0,\mathbb{E}_1)=(t,t^2),\\
    \xi^{\mu}&:=(\xi^0,\xi^1)=\left(\frac{i\langle t\rangle}{\sigma^2},-\frac{1}{2\sigma^2}\right),\\
    \phi(\xi^{\nu})&:=\frac{1}{2}\ln\left(-\frac{\pi}{\xi^1}\right)-\frac{(\xi^0)^2}{4\xi^1}=\ln\left(\sqrt{2\pi}\sigma\right)+\frac{\langle t\rangle^2}{2\sigma^2},
\end{align}
\end{subequations}
which defines the cumulant distribution as
\begin{align}
    \varrho:=-\ln(\rho(\xi))&\equiv\phi(\xi^{\nu})-\xi^{\mu} \mathbb{E}_{\mu}(\mathbb{X}^{\nu}) \notag\\
    &=\ln(\sqrt{2\pi}\sigma)+\frac{t^2}{2\sigma^2}-\frac{it\langle t\rangle}{\sigma^2}-\frac{\langle t\rangle^2}{2\sigma^2}.
\end{align}
Notice that $\mathbb{E}_{\mu}(\mathbb{X}^{\nu})$ is a function in $t$, which is one of the components $x^{\mu}$ of the classical spacetime, i.e. the stochastic variables $\mathbb{X}^{\nu}$ can be parameterized generally by the classical $x^{\mu}$ components of the spacetime as we emphasized before in the beginning of subsection (\ref{FisherKLdiv}) when we mentioned the coarse--graining the properties of the manifold. Also, $\xi^{\mu}$ variables are functions in the averages and the standard deviations of $ x^{\nu}$, i.e. $\xi^{\mu}$ are functions in statistical variables, which means we can suppress the variable $\xi$ in the metric such that $\mathpzc{G}_{\mu\nu}\left( \xi,\mathbb{X}\right)\equiv\mathpzc{G}_{\mu\nu}\left(\mathbb{X}\right)$. Moreover, defining the exponential family of probability distributions to be either complex, like how we have just done, or a real, as in Ref. \cite{Matsueda:2013saa}, will not change the definition of the cumulant probability distribution after applying the $(-\ln)$ function on Eq. (\ref{ExpFam}). Therefore, the results in the previous subsection hold correct in both approaches, where our approach is suitable for dS spaces while Ref. \cite{Matsueda:2013saa} fits for AdS. Constructing arbitrary $D$--dimensional pseudo--Riemannian information manifold from a classical spacetime is left as an exercise to the reader.\footnote{This can be obtained in a similar fashion of how AdS is obtained from complexifying dimensions of CFT, see Eq. (9--12) in Ref. \cite{Roberts:2014ifa}, or see the $(u,v)$ \emph{thermal} coordinates in Ref. \cite{Bagchi:2021qfe}.}

\section{Entropy of the Information Manifold}
\label{EntropyManifold}
As we have seen in section (\ref{EntropyPreservedInfo}) for non dissipative systems, the time rate change of the entropy is directly related to the content of information expressed in Liouville--von Neumann equation. But in the previous section we found that there is a correspondence between the spacetime and the statistical information manifold, and this correspondence makes the spacetime metric corresponds to Fisher metric even when we compare the \textit{Lorentzian} signature of both \cite{Calmet:2004a,Calmet:2004b,Caticha:2019ckh}. As the spatial entropic area $A$ is part of the spacetime, we can construct an area $\mathbb{A}$ in the information manifold that corresponds to the $A$, and the spatial metric of that information area exists in light of our discussion in subsection (\ref{Euclidstruct}). More about the area $\mathbb{A}$ of such \emph{blurred spatial space from quantum entanglement contours} can be found in Ref. \cite{Guo:2021fax,Clingman:2015lxa}. Then, it is now clear that $\mathbb{A}$ could be obtained also after complexifying the exponential distribution families.\footnote{$\mathbb{A}$ could be developed for AdS as in Ref. \cite{Matsueda:2015voa}.} And if we accept the existence of such correspondence between $A$ and $\mathbb{A}$, then we can find a equation for the spatial--like expansion rate in the information manifold \fancy{$\theta$} that corresponds to $\theta$ as mentioned before in section (\ref{EntropyPreservedInfo}), i.e.
\begin{equation}
        \fancy{$\theta$}=\frac{1}{\mathbb{A}}\frac{d}{dt}\left(\mathbb{A}\right),
\end{equation}
where, in comparison to Eq. (\ref{theta.gamma}) and Eq. (\ref{spatialFisher}), the spatial--like expansion rate in the information manifold \fancy{$\theta$} is defined using the determinant of the averaged spatial components of the Fisher metric \fancy{$\gamma$}\ = \ det(\fancy{$\gamma$}$_{ij})=\text{det}(\langle\mathfrak{g}_{ij}\rangle)$, i.e.
\begin{equation}
        \fancy{$\theta$}=\frac{1}{\sq{\fancy{$\gamma$}}}\frac{d}{dt}\left(\sq{\fancy{$\gamma$}}\right).
\end{equation}
Since the exponential family is assumed to be either real or complex, the Hamiltonian is allowed to be non--Hermitian with non-unitary transformation, and it is expected to be a stochastic Hamiltonian in comparison with regular Hamiltonians in regular phase spaces. Thus, the Hamiltonian should be modified to obey Lindblad master equation \cite{Lindblad:1975ef}, which is the most general form of Liouville equation, and we comment on this in the discussion. Then for a stochastic Hamiltonian $\mathbb{H}$ as a function in $\mathbb{X}$ and its conjugate momentum, Eq. (\ref{fine-grained}) in the information manifold becomes
\begin{equation}
    i\frac{1}{4\mathbb{G}_4}\fancy{$\theta$}\mathbb{A}=\Tr\bigg[i \hbar \frac{d \rho}{d t} \ln \rho+ [\mathbb{H},\rho]\bigg].
\end{equation}
And for dissipative systems, Eq. (\ref{result}) becomes
\begin{equation}\label{WDT}
    -i\hbar \frac{d S_{\text{gen}}}{d t}=-i\frac{1}{4\mathbb{G}_4}\fancy{$\theta$}\mathbb{A}+\Tr\bigg[i \hbar \frac{d \rho}{d t} \ln \rho+ [\mathbb{H},\rho]_{\text{Lb}}\bigg],
\end{equation}
which is the entropy of the black hole in the information manifold with no classical components from the spacetime itself, just information geometry. Based on Lindblad master equation, the $[\mathbb{H},\rho]_{\text{Lb}}$ term is the time evolution of the density under the influence of the interaction Hamiltonian $\mathbb{H}_i$ in an open dissipative system \cite{2020AIPA}
\begin{equation}
    \frac{d\rho(t)}{dt}\equiv[\mathbb{H},\rho]_{\text{Lb}}:=\frac{1}{i\hbar}[\mathbb{H}_i(t),\rho(0)]-\frac{1}{\hbar^2}\int^t_0 dt'[\mathbb{H}_i(t),[\mathbb{H}_i(t'),\rho(t)]].
\end{equation}

The question now is: what is the meaning of the area $\mathbb{A}$ in the information manifold, and how is it precisely related to the spatial extremal area $A$ in the spacetime? The answer lies in the definition of extremal surface in dS space, which corresponds to the so--called \emph{pseudo--entropy} \cite{Doi:2022iyj,Doi:2023zaf}. Such dS extremal surface \cite{Narayan:2015vda,Narayan:2015oka} could be compared with that RT surface in the AdS space. In order to get the extremal surfaces in dS space, we must complexify time such that average speed on the timelike surfaces, defined by the ratio of the shortest spatial angular length $l$ of a dS space with a radius $R_{dS}$ to the shortest time $\epsilon$, becomes the determining factor of the size of the extermal surface, i.e.

\begin{equation}
    A_{dS}=-\pi R^2_{dS}\left(\frac{l}{c\epsilon}-1\right).
\end{equation}
The last equation shows that for some cases, $l/\epsilon>1$, the area could be negative,\footnote{The area could be imaginary for dS spaces with different dimensions, see Ref. \cite{Narayan:2015oka}.} which would lead to negative or even complex valued entropy! Before the last comment gets ``frown upon'', this could be necessary to avoid the disappearance of the spatial surfaces of dS Rindler wedge \cite{Miyaji:2015yva}. This might not be well--appreciated as it says there could be non--unitary states in CFT, i.e, the corresponding Hamiltonian $\mathbb{H}$ could be non--Hermitian. Very recently, it is proved that the non--Hermiticity stems from the fact that the non--unitary CFT, dual to dS, lives on a space--like surface and the time coordinate emerging from an Euclidean CFT \cite{Narayan:2022afv} related to the previously mentioned blurred space. When we shift to the language of information manifolds, we surprisingly find some ``untimely meditations'' about the necessity of complexfying the spacetime and probability distributions so that we get a Fisher metric as an averaged metric over spacetime fluctuations with a Lorentzian signature \cite{Calmet:2004a}. Consequently, claiming the necessity of non--Hermitian Hamiltonian corresponding to the Einstein--Hilbert formulation of GR in information manifold suggests that we can study the dynamics of a Wheeler-deWitt Hamiltonian \cite{DeWitt:1967yk,wheeler1968battelle} described by the spatial metric \fancy{$\gamma$} on information manifold. An example of the pseudo--Hermitian Wheeler--deWitt Hamiltonian is discussed in details in \cite{Mostafazadeh:2001jk}, and detailed calculations of such Hamiltonian in information manifold could be followed from Ref. \cite{Carroll:2007zza,Carroll2008,Vaz:1999nt} but they are left for a future study.

One last thing to be said about the RT formula of pseudo--entropy. By comparing the approach\footnote{See Ref. \cite{Narayan:2015oka} for the analysis of RT formula in $2D$ information manifolds.} followed in \cite{Matsueda:2014yza}, the RT formula in $4D$ dS becomes
\begin{equation}
    S_{\text{gen.RT}}=\frac{A_{dS}}{4G_4}=-\frac{\pi R^2_{dS}}{4G_4}\ln(\frac{l}{c\epsilon})\sim-\frac{\pi R^2_{dS}}{4G_4}\left(\frac{l}{c\epsilon}-1\right),
\end{equation}
which is the formula of the entropy of black hole when the horizon coincides with the RT extremal surface. And to get the expression of the entropy of the extremal surface in information manifold, simply replace $A\to\mathbb{A}$ and $G_4\to\mathbb{G}_4$, where $G_4$ is the gravitational constant in $4D$ spacetime as given in Eq. (\ref{G4infomanifold}). For more details on the leading divergent term in the previous equation and its relation to the holographic entanglement entropy, see Ref. \cite{Sato:2016oaw}.

\section{Discussions and Conclusions}
\label{conclusion}

As we have seen, our study suggests reducing the geometrical properties, including spacetime itself, to an information geometry language in a way that could evolve the insight on the deep connection between physics and information. In this work, we studied in details the coarse--grained/fined--grained entropy of the black hole that obeys the second law of thermodynamics.  We analyzed the entropy--area law corrected by von Neumann entropy of the quantum matter outside its event horizon in order to obey second law of thermodynamics and to preserve information. We constructed the corresponding form of this corrected entropy--area law in quantum information geometric language. Consequently, a corresponding spacetime emerges from the quantum information. We discussed the link between Wald and Jacobson approaches of thermodynamic/gravity correspondence and Fisher pseudo--Riemannian metric of information manifold that guarantees extending the geometric interpretation to any quantum theory. We formulated Einstein's field equations in information geometry forms, and we obtained a modified Ricci tensor that helped constructing a Lagrangian of such theory. Also, we used the modified Ricci tensor to introduce the reduced Einstein tensor $\Tilde{\mathbb{G}}_{\mu\nu}$, which is directly related to the energy--momentum tensor in such manifold. The formulated Einstein's field equations led into two interesting outcomes stemming fundamentally from information geometry. The first result is finding an informatic origin of a positive cosmological constant that is founded on Fisher metric. This cosmological constant resembles those found in Lovelock's theories in de Sitter background due to complexifying time and the Gaussian exponential families of probability distributions. The second result is a time varying gravitational constant that resembles the idea of Dirac's large number hypothesis and predicts varying of  gravitational constant based on simple analysis of nature constants. We extended our analysis into the information manifold and wrote down a dynamical equation for the entropy in an information manifold using Liouville--von Neumann equation for the quantum system Hamiltonian. According to our results, the Hamiltonian in the information manifold is allowed to be non--Hermitian. The resulting dynamical equation provides a clue to a direction that could ameliorate the problem of time.

It is worth noting that relating Jacobson endeavors to information geometry requires considering \textit{non--equilibrium thermodynamics of spacetime} and its associated \textit{dissipative gravity} \cite{Eling:2006aw,Chirco:2009dc}. This relation comes from the coarse--graining process and the suggested quantum thermodynamical origin of spacetime. To achieve fine--graining, the observational entropy will match von Neumann entropy after several consecutive coarse--graining processes \cite{Safranek}. And for finite--dimensional systems like spacetime itself, as $x^{\mu}$ are obviously finite, the observational entropy can be expressed as \textit{relative entropy}. Moreover, a non--local heat fluxes is proven to take place in non--equilibrium thermodynamical gravitational systems, for both Einstein's general relativity and the scalar--tensor modifications, due to their dissipative characters \cite{Chirco:2009dc}. When applied to Rindler spacetime, The thermal character of such heat flux extends the vacuum thermal state to include all possible Rindler wedges not just that of a single observer. So, an accelerated observer would access information, due to entanglement entropy, on spacelike slices. Consequently, the restricting the neighborhood of the Rindler wedge origin, as spacelike slice around it, determines the expansion coefficient. As the time rate change of the expansion coefficient is related to the dissipative energy coupled to the bulk and shear viscosity, then it relates the entropy, both internal at the irreversible level and exchanged at the reversible level, to the Equivalence Principle in such dissipative systems. Moreover, the time rate change of the expansion coefficient states a universal relation between viscosity and internal viscous part of the entropy density as found in the AdS/CFT \cite{Maldacena:1997re}.

We know that the collapse of a pure state results in a mixed state, which is a process the unitary transformations cannot achieve in irreversible processes due to the problem of preserving the norm of the wave functions. Moreover, there is no such \textit{realistic} quantum system that could be described by a pure state \cite{Zurek:1982ii,Zurek:1981xq}. So, we are left with non--unitary transformations. As the quantum time--reversible processes governed by Schrodinger evolution equation are always unitary, then there is an obvious contradiction in assuming those processes to comprise the non--reversible macro physics such as the second law of thermodynamics. There is a long debate on the optimal epistemic and/or ontological way to resolve this contradiction, it is usually discussed under the umbrella of wave function jump and dissipative systems, see Ref. \cite{allori2020wave} for more details. One of the remarkable suggestions is that the microscopic variables indeed evolve according to non--unitary processes \cite{albert2001time}. So, based on the non--unitary irreversible von Neumann’s \textit{measurement transitions} that render mixed states \cite{von1955mathematical}, Ref. \cite{e19030106} provides an account on assuming that wave function collapse takes place spontaneously and randomly in space and time. We know that von Neumann entropy is a function in basis--independent density operators, meanwhile the \textit{reversible} Shannon entropy, related to Gibbs and Boltzmann ones, is a function in the probability density matrices. If von Neumann and Shannon entropies are related, then there must be some stochastic variables that correspond to \textit{coarse--graining} Liouville equation. Classically this is associated with a loss in information, which is discussed in Balian \emph{et al.} endeavor \cite{Balian:1986jrj}. At the quantum level, this is associated with a loss in phase coherence in quantum states. This is suggested to happen in a non--unitary transformation from a pure state to mixed one \cite{e19030106}. This guarantees the validity of the second law of thermodynamics despite having Liouville equation to be governed by phase--space Hamiltonians. However, this does not demand making change to the Schrodinger equation; rather it says that Liouville equation is physically broken at the quantum level. The price could be replacing the regular Liouville equation with the more general Lindblad master equation of open quantum systems \cite{Lindblad:1975ef}, which allows the corresponding Hamiltonian to have non--Hermitian parts \cite{2020AIPA}.
These non--unitary quantum processes are found in many high energy systems, such as CFTs with zero and negative central charges that exhibit entanglement entropy \cite{Bianchini:2014uta, Bianchini:2015uea}, in condensed matter systems \cite{PhysRevLett.119.040601}, and even in quantum electronics \cite{Mannhart_2021}.

A final comment on complexifying time related to what is mentioned in Isham’s report on the problem of time \cite{Isham:1992ms}. In the example, if we focus only on the properties of an open quantum system, then the relevant state becomes that of the reduced density matrix obtained by summing over--in another word by tracing out--the states of the surroundings. If the states of the surroundings are approximately orthogonal--Balian \emph{et al.} assume that too—then the density of quantum system exhibits decoherence. This is guaranteed to be true as the reduced density matrix is proved to be governed in different examples of Lindblad master equation \cite{Schlosshauer:2019ewh} such as spatial decoherence \cite{1985ZPhyB} and quantum Brownian motion \cite{CALDEIRA1983587}. Isham emphasizes that the inability to find a satisfactory unitary Hamiltonian for Wheeler--de Witt equation should not be considered necessarily as a disaster. Rather, this ``\textit{might reflect something of genuine physical significance}’’. For example, in Hartle and Hawking \cite{Hartle:1983ai} and Vilenkin \cite{Vilenkin:1987kf} endeavors, time becomes \textit{complex} due to the \textit{non--unitary} evolution.\\

\noindent\rule{16.1cm}{0.8pt}\\

\noindent {\Large\bf Aknowledgement}

\noindent The author would like to thank Ahmed Farag Ali for for discussions and comments during the preparation of this work.\\

\noindent\rule{16.1cm}{0.8pt}

%\newpage
\section{Appendix}\label{append}
This appendix is dedicated for the detailed calculations of the connection terms and their derivatives in Einstein equation in information manifold.\\

If we substitute Eq. (\ref{Christ}) in the derivative terms of Eq. (\ref{RicTen}) we get
\begin{align}\label{parG-parG}
    \medmath{
    \partial_{\alpha}\mathbb{\Gamma}^{\alpha}_{\mu\nu}-\partial_{\nu}\mathbb{\Gamma}^{\alpha}_{\mu\alpha}}&\medmath{=
    \mathbb{g}^{\alpha\beta}\partial_{\alpha}\left[\langle\partial_{\mu}\partial_{\nu}\varrho\partial_{\beta}\varrho\rangle-\frac{1}{2}\langle\partial_{\mu}\varrho\partial_{\nu}\varrho\partial_{\beta}\varrho\rangle\right]
    -\mathbb{g}^{\alpha\beta}\partial_{\mu}\left[\langle\partial_{\alpha}\partial_{\nu}\varrho\partial_{\beta}\varrho\rangle-\frac{1}{2}\langle\partial_{\alpha}\varrho\partial_{\nu}\varrho\partial_{\beta}\varrho\rangle\right]} \notag\\
    &\medmath{+\partial_{\alpha}(\mathbb{g}^{\alpha\beta})\left[\langle\partial_{\mu}\partial_{\nu}\varrho\partial_{\beta}\varrho\rangle-\frac{1}{2}\langle\partial_{\mu}\varrho\partial_{\nu}\varrho\partial_{\beta}\varrho\rangle\right]
    -\partial_{\mu}(\mathbb{g}^{\alpha\beta})\left[\langle\partial_{\alpha}\partial_{\nu}\varrho\partial_{\beta}\varrho\rangle-\frac{1}{2}\langle\partial_{\alpha}\varrho\partial_{\nu}\varrho\partial_{\beta}\varrho\rangle\right]} \notag\\
    \tag{A1}
\end{align}
Also, if we substitute Eq. (\ref{Christ}) in the multiplicative terms of Eq. (\ref{RicTen}) we get
\begin{align}\label{GG-GG}
    \medmath{\mathbb{\Gamma}^{\alpha}_{\beta\alpha}\mathbb{\Gamma}^{\beta}_{\mu\nu}-\mathbb{\Gamma}^{\alpha}_{\beta\nu}\mathbb{\Gamma}^{\beta}_{\mu\alpha}=\mathbb{g}^{\alpha\kappa}\mathbb{g}^{\beta\lambda}\Bigg\{ }&\medmath{\left[\langle\partial_{\alpha}\partial_{\beta}\varrho\partial_{\kappa}\varrho\rangle-\frac{1}{2}\langle\partial_{\alpha}\varrho\partial_{\beta}\varrho\partial_{\kappa}\varrho\rangle\right]\times
    \left[\langle\partial_{\mu}\partial_{\nu}\varrho\partial_{\lambda}\varrho\rangle-\frac{1}{2}\langle\partial_{\mu}\varrho\partial_{\nu}\varrho\partial_{\lambda}\varrho\rangle\right]} \notag\\
    -&\medmath{\left[\langle\partial_{\nu}\partial_{\beta}\varrho\partial_{\kappa}\varrho\rangle-\frac{1}{2}\langle\partial_{\nu}\varrho\partial_{\beta}\varrho\partial_{\kappa}\varrho\rangle\right]\times
    \left[\langle\partial_{\mu}\partial_{\alpha}\varrho\partial_{\lambda}\varrho\rangle-\frac{1}{2}\langle\partial_{\mu}\varrho\partial_{\alpha}\varrho\partial_{\lambda}\varrho\rangle\right]\Bigg\} }\tag{A2}
\end{align}
We distribute the outer derivative applied on first line in Eq. (\ref{parG-parG}), together with exploiting the properties in Eq. (\ref{partvarrho3}), to get
\begin{align}\label{1stline-parG-parG}
    &\medmath{\mathbb{g}^{\alpha\beta}\partial_{\alpha}\left[\langle\partial_{\mu}\partial_{\nu}\varrho\partial_{\beta}\varrho\rangle-\frac{1}{2}\langle\partial_{\mu}\varrho\partial_{\nu}\varrho\partial_{\beta}\varrho\rangle\right]
    -\mathbb{g}^{\alpha\beta}\partial_{\mu}\left[\langle\partial_{\alpha}\partial_{\nu}\varrho\partial_{\beta}\varrho\rangle-\frac{1}{2}\langle\partial_{\alpha}\varrho\partial_{\nu}\varrho\partial_{\beta}\varrho\rangle\right]=} \notag\\
    &\medmath{\mathbb{g}^{\alpha\beta}\Bigg\{\langle\partial_{\mu}\partial_{\nu}\varrho\partial_{\alpha}\partial_{\beta}\varrho\rangle-\langle\partial_{\alpha}\varrho\partial_{\mu}\partial_{\nu}\varrho\partial_{\beta}\varrho\rangle-\frac{1}{2}\Big[\langle\partial_{\mu}\partial_{\alpha}\varrho\partial_{\nu}\varrho\partial_{\beta}\varrho\rangle+\langle\partial_{\mu}\varrho\partial_{\alpha}\partial_{\nu}\varrho\partial_{\beta}\varrho\rangle+\langle\partial_{\mu}\varrho\partial_{\nu}\varrho\partial_{\alpha}\partial_{\beta}\varrho\rangle\Big]} \notag\\
    &\medmath{-\langle\partial_{\alpha}\partial_{\nu}\varrho\partial_{\mu}\partial_{\beta}\varrho\rangle+\langle\partial_{\mu}\varrho\partial_{\alpha}\partial_{\nu}\varrho\partial_{\beta}\varrho\rangle+\frac{1}{2}\Big[\langle\partial_{\alpha}\partial_{\mu}\varrho\partial_{\nu}\varrho\partial_{\beta}\varrho\rangle+\langle\partial_{\alpha}\varrho\partial_{\mu}\partial_{\nu}\varrho\partial_{\beta}\varrho\rangle+\langle\partial_{\alpha}\varrho\partial_{\nu}\varrho\partial_{\mu}\partial_{\beta}\varrho\rangle\Big]\Bigg\}=} \notag\\
    &\medmath{\mathbb{g}^{\alpha\beta}\Big[\langle\partial_{\mu}\partial_{\nu}\varrho\partial_{\alpha}\partial_{\beta}\varrho\rangle+\frac{1}{2}\langle\partial_{\mu}\varrho\partial_{\alpha}\partial_{\nu}\varrho\partial_{\beta}\varrho\rangle+\frac{1}{2}\langle\partial_{\mu}\varrho\partial_{\beta}\partial_{\nu}\varrho\partial_{\alpha}\varrho\rangle} \notag\\
    &\hspace*{5.9cm}\medmath{-\langle\partial_{\mu}\partial_{\beta}\varrho\partial_{\nu}\partial_{\alpha}\varrho\rangle-\frac{1}{2}\langle\partial_{\alpha}\varrho\partial_{\mu}\partial_{\nu}\varrho\partial_{\beta}\varrho\rangle-\frac{1}{2}\langle\partial_{\mu}\varrho\partial_{\alpha}\partial_{\beta}\varrho\partial_{\nu}\varrho\rangle\Big]} \tag{A3}
\end{align}
We contract Eq. (\ref{1stline-parG-parG}) using $\mathbb{g}_{\mu\nu}$ to get
\begin{align}\label{contracted-1stline-parG-parG}
    &\medmath{\mathbb{g}^{\alpha\beta}\partial_{\alpha}\left[\langle\partial_{\kappa}\partial_{\kappa}\varrho\partial_{\beta}\varrho\rangle-\frac{1}{2}\langle\partial_{\kappa}\varrho\partial_{\kappa}\varrho\partial_{\beta}\varrho\rangle\right]
    -\mathbb{g}^{\alpha\beta}\partial_{\kappa}\left[\langle\partial_{\alpha}\partial_{\kappa}\varrho\partial_{\beta}\varrho\rangle-\frac{1}{2}\langle\partial_{\alpha}\varrho\partial_{\kappa}\varrho\partial_{\beta}\varrho\rangle\right]=} \notag\\
    &\medmath{\mathbb{g}^{\alpha\beta}\Big[\langle\partial_{\kappa}\partial_{\kappa}\varrho\partial_{\alpha}\partial_{\beta}\varrho\rangle+\frac{1}{2}\langle\partial_{\kappa}\varrho\partial_{\alpha}\partial_{\kappa}\varrho\partial_{\beta}\varrho\rangle+\frac{1}{2}\langle\partial_{\kappa}\varrho\partial_{\beta}\partial_{\kappa}\varrho\partial_{\alpha}\varrho\rangle} \notag\\
    &\hspace*{5.9cm}\medmath{-\langle\partial_{\kappa}\partial_{\beta}\varrho\partial_{\kappa}\partial_{\alpha}\varrho\rangle-\frac{1}{2}\langle\partial_{\alpha}\varrho\partial_{\kappa}\partial_{\kappa}\varrho\partial_{\beta}\varrho\rangle-\frac{1}{2}\langle\partial_{\kappa}\varrho\partial_{\alpha}\partial_{\beta}\varrho\partial_{\kappa}\varrho\rangle
    \Big]} \tag{A4}
\end{align}
Now, we multiply Eq. (\ref{contracted-1stline-parG-parG}) by $-\frac{1}{2}\mathbb{g}_{\mu\nu}$ then pick terms from the result of such multiplication that match with the terms in Eq. (\ref{1stline-parG-parG}) to show that Eq. (\ref{parG-parG}), i.e. the derivative parts in Eq. (\ref{RicTen}), corresponds to the cosmological coupling constant term in Lovelock theory of gravity \cite{Lovelock:1971yv}. The middle steps are obtained using the definition of the $\mathpzc{G}_{\mu\nu}$ in Eq. (\ref{Hessian}) and the definition of $\mathbb{g}_{\mu\nu}$ in Eq. (\ref{Hessianrho}). The $\mathbb{g}^{\mu\nu}\mathbb{g}_{\mu\nu}=\mathpzc{G}^{\mu\nu}\mathpzc{G}_{\mu\nu}=D$, where $D$ is the dimension of the manifold. Collect the fourth term in the last two lines of Eq. (\ref{1stline-parG-parG}) with $-\frac{1}{2}\mathbb{g}_{\mu\nu}\times$the fourth term in the last two lines of Eq. (\ref{contracted-1stline-parG-parG}) to get
\begin{equation}\label{-(1-D/2)}
    -\mathbb{g}^{\alpha\beta}\langle\partial_{\mu}\partial_{\alpha}\varrho\partial_{\nu}\partial_{\beta}\varrho-\frac{1}{2}\mathbb{g}_{\mu\nu}\partial_{\kappa}\partial_{\alpha}\varrho\partial_{\kappa}\partial_{\beta}\varrho\rangle=-(1-D/2)\mathbb{g}_{\mu\nu} \tag{A5}
\end{equation}
Collect the second and the third terms in the last two lines of Eq. (\ref{1stline-parG-parG}) with $-\frac{1}{2}\mathbb{g}_{\mu\nu}\times$the second and third terms in the last two lines of Eq. (\ref{contracted-1stline-parG-parG}) to get
\begin{equation}\label{(1-D/2)}
    \frac{1}{2}\mathbb{g}^{\alpha\beta}\langle\partial_{\alpha}\partial_{\nu}\varrho\partial_{\beta}\partial_{\mu}\varrho+\partial_{\alpha}\partial_{\mu}\varrho\partial_{\beta}\partial_{\nu}\varrho-\mathbb{g}_{\mu\nu}\partial_{\kappa}\partial_{\alpha}\varrho\partial_{\kappa}\partial_{\beta}\varrho\rangle=(1-D/2)\mathbb{g}_{\mu\nu} \tag{A6}
\end{equation}
It is obvious that the last two results, Eq. (\ref{-(1-D/2)}) and Eq. (\ref{(1-D/2)}), cancel each other. Collect the last term in the last two lines of Eq. (\ref{1stline-parG-parG}) with $-\frac{1}{2}\mathbb{g}_{\mu\nu}\times$the last term in the last two lines of Eq. (\ref{contracted-1stline-parG-parG}) to get
\begin{equation}\label{-(D/2-D^2/4)}
-\frac{1}{2}\mathbb{g}^{\alpha\beta}\langle\partial_{\alpha}\partial_{\beta}\varrho(\partial_{\mu}\partial_{\nu}\varrho-\frac{1}{2}\mathbb{g}_{\mu\nu}\partial_{\kappa}\partial_{\kappa}\varrho)\rangle=-(D/2-D^2/4)\mathbb{g}_{\mu\nu} \tag{A7}
\end{equation}
Collect the first term in the last two lines of Eq. (\ref{1stline-parG-parG}) with $-\frac{1}{2}\mathbb{g}_{\mu\nu}\times$the first term in the last two lines of Eq. (\ref{contracted-1stline-parG-parG}) to get
\begin{equation}\label{(D/2-D^2/4)}
    \frac{1}{2}\mathbb{g}^{\alpha\beta}\langle\partial_{\alpha}\partial_{\beta}\varrho(\partial_{\mu}\partial_{\nu}\varrho-\frac{1}{2}\mathbb{g}_{\mu\nu}\partial_{\kappa}\partial_{\kappa}\varrho)\rangle=(D/2-D^2/4)\mathbb{g}_{\mu\nu} \tag{A8}
\end{equation}
It is obvious that the last two results, Eq. (\ref{-(D/2-D^2/4)}) and Eq. (\ref{(D/2-D^2/4)}), cancel each other. We are left with the fifth term in the last two lines of Eq. (\ref{1stline-parG-parG}) and the $-\frac{1}{2}\mathbb{g}_{\mu\nu}\times$fifth term in the last two lines of Eq. (\ref{contracted-1stline-parG-parG}). We combine both to get
\begin{equation}\label{CosmInfoD}
    -\frac{1}{2}\mathbb{g}^{\alpha\beta}\langle\partial_{\alpha}\varrho\partial_{\beta}\varrho(\partial_{\mu}\partial_{\nu}\varrho-\frac{1}{2}\mathbb{g}_{\mu\nu}\partial_{\kappa}\partial_{\kappa}\varrho)\rangle=-(D/2-D^2/4)\mathbb{g}_{\mu\nu} \tag{A9}
\end{equation}
Therefore, Eq. (\ref{CosmInfoD}) is the only part that contributes to Eq. (\ref{parG-parG})
\begin{equation}\label{Lambda(D/2-1)append}
    \Big(\partial_{\alpha}\mathbb{\Gamma}^{\alpha}_{\mu\nu}-\partial_{\nu}\mathbb{\Gamma}^{\alpha}_{\mu\alpha}\Big)-\frac{1}{2}\mathbb{g}_{\mu\nu}\mathbb{g}^{\kappa\lambda}\Big(\partial_{\alpha}\mathbb{\Gamma}^{\alpha}_{\kappa\lambda}-\partial_{\kappa}\mathbb{\Gamma}^{\alpha}_{\lambda\alpha}\Big)=\frac{1}{2}D(\frac{D}{2}-1)\mathbb{g}_{\mu\nu}\tag{A$9^{\prime}$}
\end{equation}
where $\Lambda=D(D/2-1)$ is defined for $D\geqslant 2$ dimensional manifold.\\

For the second line in Eq. (\ref{parG-parG}), we use
\begin{align}
    \partial_{\mu}(\mathbb{g}^{\alpha\beta})&=-\mathbb{g}^{\alpha\kappa}\mathbb{g}^{\beta\lambda}\partial_{\mu} \mathbb{g}_{\kappa\lambda}=\mathbb{g}^{\alpha\kappa}\mathbb{g}^{\beta\lambda}\langle\partial_{\mu}\varrho\partial_{\kappa}\varrho\partial_{\lambda}\varrho\rangle \tag{A10}
\end{align}
so that the second line in Eq. (\ref{parG-parG}) is
\begin{align}\label{2ndlineof-parG-parG}
    \medmath{\mathbb{g}^{\alpha\kappa}\mathbb{g}^{\beta\lambda}\Bigg\{
    \langle\partial_{\alpha}\varrho\partial_{\kappa}\varrho\partial_{\lambda}\varrho\rangle\left[\langle\partial_{\mu}\partial_{\nu}\varrho\partial_{\beta}\varrho\rangle-\frac{1}{2}\langle\partial_{\mu}\varrho\partial_{\nu}\varrho\partial_{\beta}\varrho\rangle\right] 
    -\langle\partial_{\mu}\varrho\partial_{\kappa}\varrho\partial_{\lambda}\varrho\rangle\left[\langle\partial_{\alpha}\partial_{\nu}\varrho\partial_{\beta}\varrho\rangle-\frac{1}{2}\langle\partial_{\alpha}\varrho\partial_{\nu}\varrho\partial_{\beta}\varrho\rangle\right]\Bigg\}} \tag{A11}
\end{align}
Next, we expand Eq. (\ref{GG-GG}). The terms with no $1/4$ resulting from such expansion are
\begin{align}\label{NO1/4inGG-GG}
    \mathbb{g}^{\alpha\kappa}\mathbb{g}^{\beta\lambda}\Big[&
    \langle\partial_{\alpha}\partial_{\beta}\varrho\partial_{\kappa}\varrho\rangle\langle\partial_{\mu}\partial_{\nu}\varrho\partial_{\lambda}\varrho\rangle-\frac{1}{2}\langle\partial_{\alpha}\partial_{\beta}\varrho\partial_{\kappa}\varrho\rangle\langle\partial_{\mu}\varrho\partial_{\nu}\varrho\partial_{\lambda}\varrho\rangle-\frac{1}{2}\langle\partial_{\alpha}\varrho\partial_{\beta}\varrho\partial_{\kappa}\varrho\rangle\langle\partial_{\mu}\partial_{\nu}\varrho\partial_{\lambda}\varrho\rangle \notag\\
    &-\langle\partial_{\nu}\partial_{\beta}\varrho\partial_{\kappa}\varrho\rangle\langle\partial_{\mu}\partial_{\alpha}\varrho\partial_{\lambda}\varrho\rangle+\frac{1}{2}\langle\partial_{\nu}\partial_{\beta}\varrho\partial_{\kappa}\varrho\rangle\langle\partial_{\mu}\varrho\partial_{\alpha}\varrho\partial_{\lambda}\varrho\rangle+\frac{1}{2}\langle\partial_{\nu}\varrho\partial_{\beta}\varrho\partial_{\kappa}\varrho\rangle\langle\partial_{\mu}\partial_{\alpha}\varrho\partial_{\lambda}\varrho\rangle\Big] \tag{A12}
\end{align}
For the first line in Eq. (\ref{NO1/4inGG-GG}), exchange $\beta$ and $\lambda$ in the first term, change the sign in that term according to Eq. (\ref{partvarrho3}). Then, collect that term with the third term in the same line. And for the second line in Eq. (\ref{NO1/4inGG-GG}), exchange $\kappa$ and $\alpha$ in the first term, change the sign in that term according to Eq. (\ref{partvarrho3}). Then, collect that term with the third term in the same line. Also for the second and fifth term in Eq.(\ref{NO1/4inGG-GG}), change the sign in those terms according to Eq. (\ref{partvarrho3}). Thus, Eq. (\ref{NO1/4inGG-GG}) becomes
\begin{align}\label{newNO1/4inGG-GG}
    &\medmath{-\mathbb{g}^{\alpha\kappa}\mathbb{g}^{\beta\lambda}\Bigg\{
    \langle\partial_{\alpha}\varrho\partial_{\kappa}\varrho\partial_{\lambda}\varrho\rangle\left[\langle\partial_{\mu}\partial_{\nu}\varrho\partial_{\beta}\varrho\rangle-\frac{1}{2}\langle\partial_{\mu}\varrho\partial_{\nu}\varrho\partial_{\beta}\varrho\rangle\right] 
    -\langle\partial_{\mu}\varrho\partial_{\kappa}\varrho\partial_{\lambda}\varrho\rangle\left[\langle\partial_{\alpha}\partial_{\nu}\varrho\partial_{\beta}\varrho\rangle-\frac{1}{2}\langle\partial_{\alpha}\varrho\partial_{\nu}\varrho\partial_{\beta}\varrho\rangle\right]\Bigg\}} \notag\\
    &\medmath{+\frac{1}{2}\mathbb{g}^{\alpha\kappa}\mathbb{g}^{\beta\lambda}\bigg[\langle\partial_{\alpha}\varrho\partial_{\beta}\varrho\partial_{\kappa}\varrho\rangle\langle\partial_{\mu}\varrho\partial_{\nu}\varrho\partial_{\lambda}\varrho\rangle-\langle\partial_{\nu}\varrho\partial_{\beta}\varrho\partial_{\kappa}\varrho\rangle\langle\partial_{\mu}\varrho\partial_{\alpha}\varrho\partial_{\lambda}\varrho\rangle\bigg]} \tag{A13}
\end{align}
We see that the first line in Eq. (\ref{newNO1/4inGG-GG}) cancels with Eq. (\ref{2ndlineof-parG-parG}), which is obtained from the second line in Eq.(\ref{parG-parG}). Moreover, we add the second line in Eq. (\ref{newNO1/4inGG-GG}) to the terms with $1/4$ in Eq. (\ref{GG-GG}). Then, we contract the $\mu\nu$ indices of the result of such addition, multiply it with $-\frac{1}{2}\mathbb{g}_{\mu\nu}$, and add that to the original terms before the $\mu\nu$ contraction such that we introduce a new tensor
\begin{align}\label{mathcalRappend}
    \mathbb{\widetilde{R}}_{\mu\nu}&=
    \Big(\mathbb{\Gamma}^{\alpha}_{\beta\alpha}\mathbb{\Gamma}^{\beta}_{\mu\nu}-\mathbb{\Gamma}^{\alpha}_{\beta\nu}\mathbb{\Gamma}^{\beta}_{\mu\alpha}\Big)-\frac{1}{2}\mathbb{g}_{\mu\nu}\mathbb{g}^{\kappa\lambda}\Big(\mathbb{\Gamma}^{\alpha}_{\beta\alpha}\mathbb{\Gamma}^{\beta}_{\kappa\lambda}-\mathbb{\Gamma}^{\alpha}_{\beta\kappa}\mathbb{\Gamma}^{\beta}_{\lambda\alpha}\Big) \notag\\
    &=\frac{1}{4}\mathbb{g}^{\alpha\kappa}\mathbb{g}^{\beta\lambda}\Big[\langle\partial_{\alpha}\varrho\partial_{\beta}\varrho\partial_{\kappa}\varrho\rangle\langle\partial_{\mu}\varrho\partial_{\nu}\varrho\partial_{\lambda}\varrho\rangle-\langle\partial_{\nu}\varrho\partial_{\beta}\varrho\partial_{\kappa}\varrho\rangle\langle\partial_{\mu}\varrho\partial_{\alpha}\varrho\partial_{\lambda}\varrho\rangle\Big] \tag{A14}
\end{align}

Here we follow Ref. \cite{Matsueda:2013saa}. In order to relate the Fisher metric with the entropy defined as in Eq. (\ref{Qmetric}), or Eq. (\ref{contraQmetric}), we know the density means also the relative share of certain energy state $E(\mathbb{X}^{\mu})$ from the total collection of all energy states in the partition function $Z$ \cite{Amari:1994}, i.e.
\begin{equation}\label{ExpBeta}
    \rho(\xi^{\mu};\mathbb{X}^{\mu})=\exp\left[-\beta E(\mathbb{X}^{\mu})-\ln Z(\xi^{\mu})\right] \tag{A15}
\end{equation}
which reintroduces the probability distributions to the family of exponentials. Then, we can define the density generally as
\begin{equation}
    \rho(\xi^{\mu})=\exp\left[\xi^{\mu} \mathbb{E}_{\mu}(\mathbb{X}^{\nu})-\phi(\xi^{\nu})\right] \tag{A16}
\end{equation}
which is Eq. (\ref{ExpFam}). The corresponding dual density becomes
\begin{equation}\label{varrhoxE}
    \varrho=-\ln\rho=\phi(\xi^{\nu})-\xi^{\mu} \mathbb{E}_{\mu} \tag{A17}
\end{equation}
Applying the first and the second derivative with respect to $\xi^{\mu}$ on Eq. (\ref{varrhoxE}) yields
\begin{align}
    \partial_{\mu}\varrho&=\partial_{\mu}\phi-\mathbb{E_{\mu}(\mathbb{X}^{\nu})} \label{varrhophi1} \tag{A18}\\
    \partial_{\mu}\partial_{\nu}\varrho&=\partial_{\mu}\partial_{\nu}\phi \label{varrhophi2} \tag{A19}
\end{align}
In light of Eq. (\ref{Qmetric}), Eq. (\ref{contraQmetric}) and Eq. (\ref{Fisher}), the last Eq. (\ref{varrhophi1}--\ref{varrhophi2}) can be rearranged to get
\begin{equation}\label{Hessian}
    \langle\mathpzc{G}_{\mu\nu}\rangle=\langle\partial_{\mu}\partial_{\nu}\varrho\rangle=\partial_{\mu}\partial_{\nu}\phi=\mathbb{g}_{\mu\nu} \tag{A20}
\end{equation}
Since $\langle\partial_{\mu}\varrho\rangle=0$, then we apply another differentiation and use Eq. (\ref{Hessian}) to get
\begin{equation}\label{Hessianrho}
    \mathbb{g}_{\mu\nu}=\langle\partial_{\mu}\partial_{\nu}\varrho\rangle=\langle\partial_{\mu}\varrho\partial_{\nu}\varrho\rangle \tag{A21}
\end{equation}
despite that $\partial_{\mu}\partial_{\nu}\varrho\neq\partial_{\mu}\varrho\partial_{\nu}\varrho$. Moreover, since
\begin{equation}\label{partmathbbEpartphi}
    \langle\mathbb{E}_{\mu}\rangle=\partial_{\mu}  \phi~, \tag{A22}
\end{equation}
then the last three equations give
\begin{subequations}
\begin{align}
    \mathbb{g}_{\mu\nu}&=\langle\mathbb{E}_{\mu}\mathbb{E}_{\nu}\rangle-\langle\mathbb{E}_{\mu}\rangle\langle\mathbb{E}_{\nu}\rangle \tag{A23}\\
    &=\langle\mathbb{E}_{\mu}\mathbb{E}_{\nu}\rangle-\partial_{\mu}\phi\langle\mathbb{E}_{\nu}\rangle=-\langle\partial_{\mu}\varrho\mathbb{E}_{\nu}\rangle \label{gvarrhoE} \tag{A24}
\end{align}
\end{subequations}
The last three relations will help us to construct the Christoffel symbol \cite{Amari:1994, Amari:2000}, from the connections in Eq. (\ref{GGG}), and consequently the Riemann curvature tensor as functions in the density vectors as we will see in a little bit.

Now, we substitute Eq. (\ref{ShanonKL}) and Eq. (\ref{varrhophi1}) into Eq. (\ref{mathcalRappend}), then expand, we obtain
\begin{align}\label{newmathcalRappend}
    \mathbb{\widetilde{R}}_{\mu\nu}=\frac{1}{4}&\Bigg\{D\partial_{\mu}\phi\partial_{\nu}\phi-\mathbb{g}^{\beta\kappa}\partial_{\mu}\phi\langle\mathbb{E}_{\nu}\mathpzc{G}_{\beta\kappa}\rangle-\mathbb{g}^{\alpha\lambda}\partial_{\nu}\phi\langle\mathbb{E}_{\mu}\mathpzc{G}_{\alpha\lambda}\rangle \notag\\
    &+\mathbb{g}^{\alpha\kappa}\mathbb{g}^{\beta\lambda}\Big[\langle\mathbb{E}_{\mu}\mathpzc{G}_{\alpha\lambda}\rangle\langle\mathbb{E}_{\nu}\mathpzc{G}_{\beta\kappa}\rangle-\langle\partial_{\alpha}\varrho\partial_{\beta}\varrho\partial_{\kappa}\varrho\rangle\langle\mathbb{E}_{\mu}\mathbb{E}_{\nu}\partial_{\lambda}\varrho\rangle\Big] \notag\\
    &+\mathbb{g}^{\alpha\kappa}\mathbb{g}^{\beta\lambda}\langle\partial_{\alpha}\varrho\partial_{\beta}\varrho\partial_{\kappa}\varrho\rangle\Big[\partial_{\mu}\phi\langle\mathbb{E}_{\nu}\partial_{\lambda}\varrho\rangle+\partial_{\nu}\phi\langle\mathbb{E}_{\mu}\partial_{\lambda}\varrho\rangle\Big]\Bigg\} \tag{A25}
\end{align}
We focus on the last line of Eq. (\ref{newmathcalRappend}). Eq. (\ref{varrhophi1}) and Eq. (\ref{Hessian}) yield $\langle\partial_{\alpha}\varrho\partial_{\beta}\varrho\partial_{\kappa}\varrho\rangle=\langle\partial_{\beta}\varrho\mathpzc{G}_{\alpha\kappa}\rangle=\partial_{\beta}\phi \mathbb{g}_{\alpha\kappa}-\langle\mathbb{E}_{\beta}\mathpzc{G}_{\alpha\kappa}\rangle$. And
the terms $\langle\partial_{\mu}\varrho\mathbb{E}_{\nu}\rangle=-\mathbb{g}_{\mu\nu}$ as we infer from Eq. (\ref{gvarrhoE}). Then, we expand Eq. (\ref{newmathcalRappend}) to get
\begin{equation}\label{mathcalR3terms}
    \mathbb{\widetilde{R}}_{\mu\nu}=\frac{1}{4}\mathbb{g}^{\alpha\kappa}\mathbb{g}^{\beta\lambda}\Big[\langle\mathbb{E}_{\mu}\mathpzc{G_{\alpha\beta}}\rangle\langle\mathbb{E}_{\nu}\mathpzc{G_{\kappa\lambda}}\rangle-\langle\mathbb{E}_{\mu}\rangle\langle\mathbb{E}_{\nu}\rangle \mathbb{g}_{\alpha\beta}\mathbb{g}_{\kappa\lambda}-\langle\partial_{\alpha}\varrho\partial_{\beta}\varrho\partial_{\kappa}\varrho\rangle\langle\mathbb{E}_{\mu}\mathbb{E}_{\nu}\partial_{\lambda}\varrho\rangle\Big] \tag{A26}
\end{equation}
The last term in Eq. (\ref{mathcalR3terms}) is negligable as $\langle\mathbb{E}_{\mu}\mathbb{E}_{\nu}\partial_{\lambda}\varrho\rangle\sim-\langle\partial_{\lambda}(\mathbb{E}_{\mu}\mathbb{E}_{\nu})\rangle$ as Eq. (\ref{varrhophi2}) says that $\partial_{\lambda}\mathbb{E}_{\mu}=0=\mathbb{E}_{\nu}\partial_{\lambda}\mathbb{E}_{\mu}$, which means $\langle\mathbb{E}_{\mu}\partial_{\lambda}\mathbb{E}_{\nu}+\mathbb{E}_{\nu}\partial_{\lambda}\mathbb{E}_{\mu}\rangle=\langle\partial_{\lambda}(\mathbb{E}_{\mu}\mathbb{E}_{\nu})\rangle=0$. Therefore, Eq. (\ref{mathcalR3terms}) becomes
\begin{equation}\label{mathcalRfinal}
    \mathbb{\widetilde{R}}_{\mu\nu}=\frac{1}{4}\mathbb{g}^{\alpha\kappa}\mathbb{g}^{\beta\lambda}\Big[\langle\mathbb{E}_{\mu}\mathpzc{G_{\alpha\beta}}\rangle\langle\mathbb{E}_{\nu}\mathpzc{G_{\kappa\lambda}}\rangle-\langle\mathbb{E}_{\mu}\rangle\langle\mathbb{E}_{\nu}\rangle \mathbb{g}_{\alpha\beta}\mathbb{g}_{\kappa\lambda}\Big] \tag{A27}
\end{equation}
As we defined the stochastic variables $\mathbb{X}^{\mu}\equiv \mathbb{X}^{\mu}(\langle x^{\nu}\rangle,\sigma_{x^{\nu}})$ in the beginning of subsection (\ref{FisherKLdiv}), the same can be done for the stochastic metric $\mathpzc{G}_{\mu\nu}( \xi,\mathbb{X})$ as we expand it around the $\langle x^{\mu}\rangle$ while we keep $\sigma_{x^{\mu}}$ as it is. So, $\mathbb{X}\equiv \mathbb{X}^{\mu}(\langle x\rangle)$, And the metric becomes
\begin{equation}\label{expandedG}
    \mathpzc{G}_{\mu\nu}(\mathbb{X})=\mathpzc{G}_{\mu\nu}(\langle x\rangle)+\dot{\mathpzc{G}}_{\mu\nu}(\langle x\rangle)\Big( \mathbb{X}-\langle x\rangle\Big)+\frac{1}{2}\ddot{\mathpzc{G}}_{\mu\nu}(\langle x\rangle)\Big( \mathbb{X}-\langle x\rangle\Big)^2+\cdots \tag{A28}
\end{equation}
where
\begin{equation}
    \dot{\mathpzc{G}}_{\mu\nu}(\langle x\rangle)=\lim\limits_{\mathbb{X}\to\langle x\rangle}\frac{\partial}{\partial \mathbb{X}}{\mathpzc{G}}_{\mu\nu}(\mathbb{X}) \tag{A29}
\end{equation}
and $\ddot{\mathpzc{G}}_{\mu\nu}(\langle x\rangle)$ is the usual second derivative of the above equation. Defining $\mathpzc{G}_{\mu\nu}$ as a function in $\langle x\rangle$ allows us get $\langle\mathpzc{G}_{\mu\nu}(\langle x\rangle)\rangle=\mathpzc{G}_{\mu\nu}(\langle x\rangle)=\mathpzc{G}_{\mu\nu}$ as averaging the average is a redundant process. Now we substitute Eq. (\ref{expandedG}) in Eq. (\ref{mathcalRfinal}), together with the help of Eq. (\ref{varrhophi1}-\ref{partmathbbEpartphi}) and the approximation $\langle\mathbb{E}_{\mu}(\mathbb{X}-\langle x\rangle)^n\rangle\sim\partial_{\mu}\phi\langle(\mathbb{X}-\langle x\rangle)^n\rangle$, to obtain
\begin{align}
    \mathbb{\widetilde{R}}_{\mu\nu}=\frac{1}{4}\mathbb{g}^{\alpha\kappa}\mathbb{g}^{\beta\lambda}\Bigg\{&\Big[\mathpzc{G}_{\alpha\beta}\mathpzc{G}_{\kappa\lambda}-\mathbb{g}_{\alpha\beta}\mathbb{g}_{\kappa\lambda}\Big] 
     +\mathpzc{G}_{\kappa\lambda}\Big[\dot{\mathpzc{G}}_{\alpha\beta}\langle\mathbb{X}-\langle x\rangle\rangle+\frac{1}{2}\ddot{\mathpzc{G}}_{\alpha\beta}\langle(\mathbb{X}-\langle x\rangle)^2\rangle+\cdots\Big] \notag\\
    & +\mathpzc{G}_{\alpha\beta}\Big[\dot{\mathpzc{G}}_{\kappa\lambda}\langle\mathbb{X}-\langle x\rangle\rangle+\frac{1}{2}\ddot{\mathpzc{G}}_{\kappa\lambda}\langle(\mathbb{X}-\langle x\rangle)^2\rangle+\cdots\Big]\Bigg\}\times\partial_{\mu}\phi\partial_{\nu}\phi \tag{A30}
\end{align} %\\

%\noindent\rule{16.1cm}{0.8pt}\\

%\noindent {\Large\bf Data Availability Statement}

%\noindent No Data associated in the manuscript.

\bibliographystyle{utcaps}
\bibliography{ref.bib}{}

\end{arabicfootnotes}
\end{document}